This paper is typeset using LaTeX with the AMS-TeX package, Version
2.1 installed.  The AMS-TeX package may be received by FTP from
e-math.ams.org or by mail for a small fee from the TeX Users Group,
P.O. Box 869, Santa Barbara, California 93102, USA.

This file consists of two parts:  commands.sty and paper.tex.
Save the appropriate pieces below to the indicated name and
LaTeX twice.
\newcommand{\Ahat}{{\hat{A}}}
\newcommand{\Atilde}{{\tilde{A}}}

\newcommand{\bfrak}{{\frak b}}

\newcommand{\C}{{\Bbb C}}
\newcommand{\Dcal}{{\cal D}}
\newcommand{\Dcalaq}{{\mathop{\Dcal}^a_q}}

\newcommand{\hatDcal}{{\hat{\Dcal}}}
\newcommand{\F}{{\Bbb F}}
\newcommand{\gfrak}{{\frak g}}
\newcommand{\gltilde}{{\widetilde{gl}}}
\newcommand{\hatgl}{{\widehat{gl}}}

\newcommand{\Lcal}{{\cal L}}
\newcommand{\Lmslambda}{L^{[ \vec m]}_{\vecs} ( \vec\lambda )}

\newcommand{\Mtilde}{\widetilde{M}}
\newcommand{\vecm}{{\vec m}}
\newcommand{\N}{{\Bbb N}}

\newcommand{\Ocal}{{\cal O}}
\newcommand{\pfrak}{{\frak p}}
\newcommand{\Pcal}{{\cal P}}
\newcommand{\Scal}{{\cal S}}
\newcommand{\sfrak}{{\frak s}}
\newcommand{\vecs}{{\vec s}}
\newcommand{\Ucal}{{\cal U}}
\newcommand{\Z}{{\Bbb Z}}

\newcommand{\ad}{{\rm ad}}
\newcommand{\Ada}{{\rm Ada}}
\newcommand{\Aut}{{\rm Aut}}
\newcommand{\End}{{\rm End}\,}
\newcommand{\Hom}{{\rm Hom}}
\newcommand{\Ker}{{\rm Ker}\,}
\newcommand{\Mat}{{\rm Mat}}
\newcommand{\ord}{{\rm ord}\,}
\newcommand{\R}{{\Bbb R}}
\newcommand{\re}{{\rm Re}\,}
\newcommand{\Res}{{\rm Res}\,}
\newcommand{\resp}{{\rm resp}\,}
\newcommand{\tr}{{\rm tr}}

\newcommand{\Vir}{{\rm Vir}\,}

\catcode`@=11
\def\simrightarrow{\mathrel{\mathpalette\@vereq\sim}} % similarity sign
\def\@vereq#1#2{\lower.5\p@\vbox{\baselineskip\z@skip\lineskip-.5\p@
    \ialign{$\m@th#1\hfil##\hfil$\crcr#2\crcr\rightarrow\crcr}}}

\def\myref#1{(\ref{#1})}

\def\@sect#1#2#3#4#5#6[#7]#8{\ifnum #2>\c@secnumdepth
     \let\@svsec\@empty\else
     \refstepcounter{#1}\edef\@svsec{\csname the#1\endcsname.\hskip 1em}\fi
     \@tempskipa #5\relax
      \ifdim \@tempskipa>\z@
        \begingroup #6\relax
          \@hangfrom{\hskip #3\relax\@svsec}{\interlinepenalty \@M #8\par}%
        \endgroup
       \csname #1mark\endcsname{#7}\addcontentsline
         {toc}{#1}{\ifnum #2>\c@secnumdepth \else
                      \protect\numberline{\csname the#1\endcsname}\fi
                    #7}\else
        \def\@svsechd{#6\hskip #3\relax  %% \relax added 2 May 90
                   \@svsec #8\csname #1mark\endcsname
                      {#7}\addcontentsline
                           {toc}{#1}{\ifnum #2>\c@secnumdepth \else
                             \protect\numberline{\csname the#1\endcsname}\fi
                       #7}}\fi
     \@xsect{#5}}

\catcode`@=\active % at signs are no longer letters

\def\qed{%      This creates a degree symbol no matter where one is.
    \relax
    \ifmmode  {\hfill\Box}
    \else  {\hfill$\Box$}
    \fi
}

\newcounter{mysubsection}
\newcommand{\mysubsection}{\setcounter{equation}{0}\stepcounter{mysubsection}%
\noindent{\bf \arabic{section}.\arabic{mysubsection}.\hspace{2em}}}

\newcommand{\mysection}[1]{\setcounter{mysubsection}{0}\section{#1}}

\newcommand{\Example}{\noindent{\bf Example.~~}}
\newcommand{\Remark}{\noindent{\bf Remark.~~}}
\newcommand{\Remarks}{\noindent{\bf Remarks.~~}}
\newcommand{\Proof}{\noindent{\bf Proof.~~}}

\newcommand{\Definition}{\noindent{\bf Definition.~~}}

\newenvironment{theorem}%
{\begin{em}{\bf Theorem.~~}}%
{\end{em}}

\newenvironment{proposition}%
{\begin{em}{\bf Proposition.~~}}%
{\end{em}}

\newenvironment{lemma}%
{\begin{em}{\bf Lemma.~~}}%
{\end{em}}

\newenvironment{corollary}%
{\begin{em}{\bf Corollary.~~}}%
{\end{em}}
\documentstyle[11pt,commands]{article}

\newcommand{\parens}[1]{\left( #1 \right)}
\newcommand{\braces}[1]{\left\{ #1 \right\}}
\newcommand{\brackets}[1]{\left[ #1 \right]}

\input amssym.def

\begin{document}
\begin{titlepage}
  \title{Quasifinite highest weight modules over the Lie algebra
  of differential operators on the circle}
  \author{V. Kac and A. Radul}
  \date{}

\end{titlepage}

\maketitle

{\bf Abstract} We classify positive energy representations with
finite degeneracies of the Lie algebra $W_{1+\infty}\/$ and
construct them in terms of representation theory of the Lie
algebra $\hatgl ( \infty R_m )\/$ of infinite matrices with
finite number of non-zero diagonals over the algebra $R_m = \C [
t ] / ( t^{m + 1} )\/$.  The unitary ones are classified as well.
Similar results are obtained for the sin-algebras.

\setcounter{section}{-1}
\mysection{Introduction}
\label{sec:intro}

\mysubsection
Recent progress in conformal field theory revealed some unusual
mathematical objects called the $W_n\/$-algebras \cite{Z}.  These
algebras turned out to be quantizations of the second
Gelfand-Dickey structure for Lax equations \cite{FL}.  The complicated
structure of these algebras is greatly simplified in the limit $n
= \infty\/$ , the limiting algebra being the Lie algebra
$\hatDcal\/$, the universal central extension of the Lie algebra
of differential operators
on the circle \cite{KP}.  (Physicists denote this Lie algebra by
$W_{1+\infty}\/$ \cite{PSR}.) The possibility to get $W_n\/$
from $\hatDcal\/$ has been studied in \cite{R}, \cite{RV}.

The main goal of the present paper is to  classify and describe
the irreducible quasifinite highest weight representation of the
Lie algebra $\hatDcal\/$.  The basic technical tool is the
analytic completion $\hatDcal^\Ocal\/$ of $\hatDcal\/$ and a
family of its homomorphisms onto the central extension of the Lie
algebra $\gltilde ( \infty , R_m)\/$ of infinite matrices with
finitely many non-zero diagonals over the ring $R_m = \C [ t ] /
( t^{m+1} )\/$.

The Lie algebra $\hatDcal\/$ may be obtained via a general
construction (explained in Section 1) as a twisted Laurent
polynomial algebra over the polynomial algebra $\C [ w ]\/$.  It
is easy to see that the only other Lie algebras obtained by this
construction from $\C [ w ]\/$ are Lie algebras $\hatDcal_q\/$,
the central extension of the Lie algebra of difference operators on the
circle.  It turns out, however, that a representation theory
similar to that of $\hatDcal\/$, may be developed for a larger
Lie algebra, the central extension $\hat\Scal_q\/$ of the Lie
algebra of pseudo-difference operators on the circle (see Section
6).  The latter Lie algebra has been studied recently by many
authors (see \cite{FFZ}, \cite{GL} and references there).

Being of a very general nature, our methods may be applied to
many other examples of infinite-dimensional Lie algebras.  Noting
that $\Dcal\/$ (resp.\ $\Scal_q\/$) is a quantization of the
Poisson Lie algebra of functions on the cylinder (resp.\
2-dimensional torus), one may expect that our approach could be
extended to the quantizations of general symplectic manifolds.

\mysubsection
Let us give here the main definitions which will be used for
various examples throughout the paper.

Consider a $\Z\/${\em-graded\/} Lie algebra over $\C\/$:
$$
\gfrak = \bigoplus_{j \in \Z} \gfrak_j, \quad
[ \gfrak_i, \gfrak_j ] \subset \gfrak_{i + j}.
$$
(We do not assume $\gfrak_i\/$ to be finite-dimensional.)
We let
$$
\gfrak_\pm = \bigoplus_{j > 0} \gfrak_{\pm j}.
$$
A subalgebra $\frak p\/$ of $\gfrak\/$ is called {\em parabolic\/} if
it contains $\gfrak_0 + \gfrak_+\/$ as a proper subalgebra.

A $\gfrak\/$-module $V\/$ is called graded if
$$
V = \bigoplus_j V_j, \quad
\gfrak_i V_j \subset V_{i + j}.
$$
A graded $\gfrak\/$-module $V\/$ is called {\em quasifinite\/} if
$$
\dim V_j < \infty\; \mbox{for all}\; j.
$$

Given $\lambda \in \gfrak^\ast_0\/$, a {\em highest weight
module\/} is a $\Z\/$-graded $\gfrak\/$-module $V ( \gfrak,
\lambda ) = \bigoplus_{j \in \Z_+} V_{-j}\/$ defined by the
following properties:
\begin{enumerate}
\renewcommand{\labelenumi}{(\roman{enumi})}
\item $V_0 = \C v_\lambda\/$, where $v_\lambda\/$ is a non-zero
  vector,

\item $h v_\lambda = \lambda ( h ) v_\lambda\/$ for $h \in
  \gfrak_0\/$,

\item $\gfrak_+ v_\lambda = 0\/$,

\item $\Ucal ( \gfrak_- ) v_\lambda = V ( \gfrak, \lambda )\/$.
\end{enumerate}
Here and further $\Ucal( \sfrak )\/$ stands for the universal
enveloping algebra of the Lie algebra $\sfrak\/$.

A non-zero vector $v \in V ( \gfrak, \lambda )\/$ is called {\em
singular\/} if $\gfrak_+ v = 0\/$.  The module $V ( \gfrak,
\lambda )\/$ is irreducible if and only if any of its singular
vectors is a multiple of $v_\lambda\/$.

The ``largest'' among the modules $V ( \gfrak, \lambda )\/$ with
a given $\lambda\/$ is the {\em Verma module\/} $M ( \gfrak,
\lambda )\/$ defined by the property that the map
$$
\varphi : \Ucal ( \gfrak_- ) \rightarrow M ( \gfrak, \lambda )
$$
given by $\varphi ( u ) = u ( v_\lambda )\/$ is a vector space
isomorphism.

Any highest weight module $V ( \gfrak, \lambda )\/$ is a quotient
of $M ( \gfrak, \lambda )\/$.  The ``smallest'' among the $V (
\gfrak, \lambda )\/$ is the irreducible module $L ( \gfrak,
\lambda )\/$ (which is a quotient of $M ( \gfrak, \lambda )\/$ by
the maximal graded submodule).

We shall write $M ( \lambda )\/$ and $L ( \lambda )\/$ in place
of $M ( \gfrak, \lambda )\/$ and $L ( \gfrak, \lambda )\/$ if no
ambiguity may arise.

\mysubsection
It is useful to note that the Verma modules can be constructed as
follows:
$$
M ( \gfrak, \lambda ) = \Ucal ( \gfrak ) \otimes_{\Ucal (
\gfrak_0 + \gfrak_+ )} \C_\lambda,
$$
where $\C_\lambda\/$ is the 1-dimensional $\gfrak_0 +
\gfrak_+\/$-module given by $h \mapsto \lambda ( h )\/$ if $h \in
\gfrak_0\/$, $\gfrak_+ \mapsto 0\/$, and the action of $\gfrak\/$
is induced by the left multiplication in $\Ucal ( \gfrak )\/$.

Now, let $\pfrak = \oplus_j \pfrak_j\/$ be a parabolic subalgebra
of $\gfrak\/$, and let $\lambda \in \gfrak^\ast_0\/$ be such that
$\lambda |_{\pfrak_0 \wedge [ p, p ]} = 0\/$.  Then the $\gfrak_0
+ \gfrak_+\/$-module $\C_\lambda\/$ extends to $\pfrak\/$ by
letting $\pfrak_j \mapsto 0\/$ for $j < 0\/$, and we may
construct the highest weight
module
$$
M ( \gfrak, \pfrak, \lambda ) =
\Ucal ( \gfrak ) \otimes_{\Ucal ( \pfrak )} \C_\lambda.
$$
It is called the {\em generalized Verma module\/}.  It may be
characterized by the property that the map $\varphi\/$ induces an
isomorphism $\Ucal ( \gfrak_- ) / \Ucal ( \pfrak \cap \gfrak_- )
\rightarrow M ( \gfrak, \pfrak, \lambda )\/$.

Note that if $\dim \gfrak_j < \infty\/$ for all $j\/$, the
$\gfrak\/$-module $L ( \lambda )\/$ for any $\lambda\/$ is
quasifinite.  If however $\dim \gfrak_j = \infty\/$, which is the
case in all of our examples, the classification of quasifinite
irreducible highest weight modules becomes a non-trivial problem.
The answer to this problem for the Lie algebra $\hatDcal\/$ is
given by Theorem 4.2.  Moreover, we give an explicit construction
of all these modules in terms of irreducible highest weight
modules over the Lie algebra $\hatgl ( \infty, R_m )\/$ (Theorems
4.5 and 4.6).

\mysubsection
Recall that an {\em anti-involution\/} of a Lie algebra
$\gfrak\/$ over $\C\/$ is an additive map $\omega : \gfrak
\rightarrow \gfrak\/$ such that
$$
\omega ( \lambda a ) = \bar{\lambda} a, \quad
\omega ( [ a, b ] ) = [ \omega ( b ) , \omega ( a ) ],
\quad \mbox{for}\quad
\lambda \in \C, \quad a, b \in \gfrak.
$$
Given a $\gfrak\/$-module $V\/$, a Hermitian form $h\/$ on $V\/$
is called {\em contravariant\/} if for any $a \in \gfrak\/$
the operators $a\/$ and $\omega ( a )\/$ are (formally) adjoint
operators on $V\/$ with respect to $h\/$.

Fix an anti-involution $\omega\/$ of the Lie algebra $\gfrak\/$
such that $\omega ( \gfrak_j ) = \gfrak_{-j}\/$.  Let $L (
\gfrak, \lambda )\/$ be an irreducible highest weight module over
$\gfrak\/$ such that $\lambda ( h ) \in \R\/$ if $\omega ( h )
= h\/$.  For $v \in L ( \gfrak, \lambda )\/$ denote by $\langle v
\rangle\/$ the coefficient of $v_\lambda\/$ in the decomposition
of $v\/$ with respect to the gradation of $L ( \gfrak, \lambda
)\/$.  Let
$$
h ( a v_\lambda, b v_\lambda ) =
\langle \omega ( a ) b v_\lambda \rangle, \qquad
a, b \in U ( \gfrak ).
$$
It is easy to show (see e.g.\ [K, Chapter 9]) that $h\/$ is the
unique contravariant form on $L ( \gfrak, \lambda )\/$ such that
$h ( v_\lambda, v_\lambda ) = 1\/$; moreover, it is
non-degenerate and $h ( L ( \gfrak, \lambda )_i, L ( \gfrak,
\lambda )_j ) = 0\/$ if $i \neq j\/$.

The $\gfrak\/$-module $L ( \gfrak, \lambda )\/$ is called {\em
unitary\/} (with respect to $\omega\/$) if the contravariant form
$h\/$ is positive definite (this is independent of the choice of
$v_\lambda \in L ( \gfrak, \lambda )_0\/$).

The classification of unitary quasifinite (irreducible) highest
weight modules over $\hatDcal\/$ is given by Theorem 5.2.

Let us note in conclusion that the classification of irreducible
quasifinite highest weight $\hatDcal\/$-modules is expressed in
terms of Bernoulli polynomials.  Is it an indication of a
connection to the Riemann-Roch theorems?

\mysubsection
One of the authors wishes to thank D.~Lebedev and
M.~Golenishcheva-Kutuzova for illuminating discussions on the sin
Lie algebra.

\mysection{Twisted Laurent polynomial algebras and associated
Lie algebras.}
\label{sec:twisted-laurent}

\mysubsection
Let $A\/$ be an associative algebra over a field $\F\/$ and let
$\sigma\/$ be an automorphism of $A\/$.  Define the {\em twisted
Laurent polynomial algebra\/} $A_\sigma [ z, z^{-1} ]\/$ over
$A\/$ in the indeterminate $z\/$ to be the vector space $\F [
z, z^{-1} ] \otimes_\F A\/$ over $\F\/$ of finite sums of the
form $\sum_{j \in \Z} z^j \otimes a_j\/$, $a_j \in A\/$, with
multiplication defined by the rule
\begin{equation}
  ( z^k \otimes a ) ( z^m \otimes b ) = z^{k + m} \otimes
  \sigma^m ( a ) b,\qquad a, b \in A, \quad k, m \in \Z.
  \label{eq:multiplication}
\end{equation}
Further on we shall often write $z^m a\/$ in place of $z^m \otimes
a\/$.

\Remarks
(a)~~Replacing $z\/$ by $z a^{-1}\/$, where $a\/$ is an
invertible element of $A\/$, corresponds to replacing $\sigma\/$
by $( A d a ) \sigma\/$, where $A d a\/$ stands for the inner
automorphism:
$$
( A d a ) b = a b a^{-1}, \qquad b \in A.
$$

(b)~~Applying an automorphism $\alpha\/$ to $A\/$ replaces
$\sigma\/$ by $\alpha^{-1} \sigma \alpha\/$.

Thus, we obtain the following proposition.

\begin{proposition}
  Twisted Laurent polynomial algebras over an associative algebra
  $A$ are parameterized by the conjugacy classes of the group
  $\Aut A / A d A$.\qed
\end{proposition}

Two automorphisms of $A\/$ whose images lie in the same conjugacy
class of $\Aut A / AdA\/$ are called {\em equivalent\/}.

\mysubsection
The algebra $A_\sigma [ z, z^{-1} ]\/$ has a canonical
$\Z\/$-gradation, called the {\em principal\/} gradation:
\begin{equation}
  A_\sigma [ z, z^{-1} ] = \bigoplus_{j \in \Z} ( z^j A ).
  \label{eq:principal}
\end{equation}

Let $\Pcal = \bigoplus_{j \in \Z} \Pcal_j\/$ be a  parabolic
subalgebra of $A_\sigma [ z, z^{-1} ]\/$ .
It is clear that $\Pcal_{-1} = z^{-1} I\/$, where
$I\/$ is a (two-sided) ideal of the algebra $A\/$.  Hence
$\Pcal\/$ contains the following {\em minimal parabolic
subalgebra\/} $\Pcal (I)\/$ associated to $I\/$:
\begin{equation}
  \Pcal ( I ) =
  \left(
    \bigoplus_{j > 0} ( z^{-1} I )^j
  \right) \bigoplus \left( \bigoplus_{j \geq 0} ( z^j A ) \right) .
\end{equation}

\Remark
Given two ideals $I\/$ and $J\/$ of $A\/$, we have the following
graded subalgebra of $A_\sigma [ z, z^{-1} ]\/$:
\begin{equation}
  \Pcal ( I, J ) =
  \left(
    \bigoplus_{j > 0} ( z^{-1} I )^j
  \right) \oplus A \oplus
  \left(
    \bigoplus_{j > 0} ( z J )^j
  \right).
  \label{eq:gradedsubalgebra}
\end{equation}

\mysubsection
We denote by $\tilde A_\sigma\/$ the algebra $A_\sigma [ z, z
^{-1} ]\/$ viewed as a Lie algebra with respect to the usual
bracket:
$$
[ f, g ]' = fg - gf.
$$

Fix a trace on the algebra $A\/$, i.e., a linear map $\tr : A
\rightarrow V\/$, where $V\/$ is a vector space over $\F\/$, such
that $\tr (ab) = \tr (ba)\/$.  Then we may construct a remarkable
central extension $\Ahat_{\sigma,\tr}\/$ of $\tilde{A}_\sigma\/$ by a
central subalgebra V:
$$
0 \rightarrow V \rightarrow \Ahat_{\sigma,\tr} \rightarrow A_\sigma
\rightarrow 0
$$
as follows.  It is straightforward to check that the formula
\begin{equation}
  \begin{array}{rcl}
    \lefteqn{\Psi_{\sigma,\tr} ( z^r a, z^s b )} \\
    & = & - \Psi_{\sigma,\tr} ( z^s b, z^r a ) =
    \left\{
      \begin{array}{cl}
        \tr
        \left( (1 + \sigma + \cdots + \sigma^{r-1} )
          \left( \sigma^{-r} (f) g \right)
        \right) & \mbox{if}\; r = -s > 0,\\
        0 & \mbox{if}\; r+s \neq 0\; \mbox{or}\; r = s = 0
      \end{array}
    \right.
  \end{array}
  \label{eq:cocycle}
\end{equation}
defines a 2-cocycle on $\tilde A_{\sigma}\/$ with values in $V\/$.
Then $\Ahat_{\sigma,tr} = \Atilde + V\/$ with $V\/$ central and the
bracket of two elements $f, g \in \Atilde \subset
\Ahat_{\sigma,\tr}\/$ is given by the usual formula:
$$
[ f, g ] = [ f, g ]' + \Psi_{\sigma,\tr} ( f, g ).
$$

\Remark
(a) Replacing $z\/$ by $z a^{-1}\/$ corresponds to replacing
$\Psi_{\sigma, \tr}\/$ by $\Psi_{(\Ada) \sigma, \tr}\/$.

(b) Applying an automorphism $\alpha\/$ to $A\/$ replaces
$\Psi_{\sigma,\tr}\/$ by $\Psi_{\alpha^{-1} \sigma \alpha, \tr \circ
\alpha}\/$.

(c) Since $\Psi_{\sigma, \tr} ( z^r, z^s ) = \tr (1) r \delta_{r,
-s}\/$, the cocycle $\Psi_{\sigma, \tr}\/$ is nontrivial if $\tr
(1) \neq 0\/$.

(d) Suppose that the map $\sigma - 1 : A \rightarrow A\/$ is
surjective and that $\tr\/$ vanishes on its kernel.  Then we have
an isomorphism $\sigma - 1 : A / \Ker ( \sigma - 1 )
\simrightarrow A\/$, and $\varphi := \tr \circ ( 1 - \sigma
)^{-1}: A \rightarrow V\/$ is a well-defined map.  We have:
$$
\Psi_{\sigma, \tr} ( z^r f, z^{-r} g) = \varphi
\left( [ z^r f, z^{-r} g ] \right),
$$
hence in this case the cocycle $\Psi_{\sigma, \tr}\/$ is trivial.

(e) Suppose that $\tr ( \sigma ( a ) ) = \tr\, a\/$, $a \in A\/$.
Then $\tr\/$ extends to a trace of the algebra $A_\sigma [ z, z
^{-1} ]\/$ by letting $\tr ( z^k a ) = \delta_{k, 0} \tr\, a\/$.

\Example
Let $A = \Mat_n \F\/$; then any automorphism of $A\/$ is
equivalent to $\sigma = 1\/$ (by Remark 1(a)).  Take the usual
trace $\tr: A \rightarrow \F\/$, then $\Ahat_{\sigma,\tr}\/$ is
isomorphic to the usual affine algebra $gl_n ( \F )^\wedge\/$.

We have the corresponding to \myref{eq:principal}
$\Z\/$-gradation:
\begin{equation}
  \Ahat_{\sigma, \tr} =
  \bigoplus_j \hat A_j, \quad \mbox{where}\;
   \hat A_j = z^j A\; \mbox{if}\;
  j \neq 0, \; \mbox{and}\;  \hat A_0 =
  A + V.
  \label{eq:corresponding}
\end{equation}
For each (two-sided) non-zero ideal $I\/$ of $A\/$ we have the
associated parabolic subalgebra of $\hat A_{\sigma, \tr}\/$
\begin{equation}
  \frak p ( I ) = \bigoplus_{j > 0} ( z^{-1} I )^j \bigoplus
  \left( A \bigoplus V \right) \bigoplus \left( \bigoplus_{j > 0} z^j A
  \right).
  \label{eq:parsub}
\end{equation}

\mysubsection
We turn now to the main examples of the Lie algebras
$\Atilde_\sigma\/$ and $\Ahat_\sigma\/$, those associated to
the polynomial algebra $A = \C [ w ]\/$ in the indeterminate
$w\/$.  We show that the Lie algebras $\Atilde_\sigma\/$ are
isomorphic to the Lie algebras of all regular differential
(resp.\ difference) operators on the punctured complex plane
$\C^\times = \C \backslash \{ 0 \}\/$, and that the Lie algebras
$\Ahat_{\sigma,\tr}\/$ are their well known central extensions.

For $q \in \C^\times\/$ define the following operator on $\C [ z,
z^{-1} ]\/$:
$$
D_q f(z) =
\left\{
  \begin{array}{c@{\quad}l}
    \displaystyle
    \frac{f (qz) - f (z)}{q - 1} & \mbox{if}\; q \neq 1,\\
    \displaystyle
    z \partial_z f ( z ) & \mbox{if}\; q = 1.
  \end{array}
\right.
$$
Denote by ${\mathop{\Dcal}^a_q}\/$ the associative algebra of all operators
on $\C [ z, z^{-1} ]\/$ of the form
$$
E = e_k (z) D^k_q + e_{k-1}(z) D^{k-1}_q + \cdots + e_0 (z
),\quad \mbox{where}\quad e_i ( z ) \in \C [ z, z^{-1} ]
$$
(the superscript $a\/$ stands for ``associative'') and let
$\Dcal_q\/$ denote the corresponding Lie algebra.

Now, any automorphism of $\C [ w ]\/$ is equivalent to
$\sigma_q\/$, $q \in \C^\times\/$, defined by
$$
\sigma_q ( w ) = q w + 1.
$$
Note that
\begin{equation}
  \sigma^n_q ( w ) = q^n w + [ n ],
  \label{eq:note}
\end{equation}
where, as usual, for $n \in \Z\/$:
$$
[ n ] = \frac{q^n - 1}{q - 1}\quad \mbox{if}\; q \neq 1\quad
\mbox{and}\; = n\; \mbox{if}\; q = 1.
$$

\begin{proposition}
(a) The linear map $\C [ w ]_{\sigma_q} [ z, z^{-1}
] \rightarrow \Dcalaq$ defined by $z^k f(w) \mapsto z^k
    f(D_q)$ is an isomorphism of associative algebras.

(b) Let $\tr: \C [ w ] \rightarrow \C$ be the
    evaluation map at $w = 0\/$.  Then the 2-cocycle
    $\Psi_{\sigma_q,\tr}\/$ on the Lie algebra $\C [ w
    ]_{\sigma_q} [ z, z^{-1} ]$ induces, via the
    above isomorphism, the following 2-cocycle on the Lie algebra
    $\Dcal_q$:
    \begin{equation}
      \Psi
      \left(
        z^m f(D_q), z^n g(D_q)
      \right) =
      \left\{
      \begin{array}{c@{\quad}l}
      \sum_{-m \leq j \leq -1}
      f \left( [j] \right) g \left( [j + m] \right) & \mbox{if}\;
      m = -n \geq 0,\\
      0 & \mbox{if}\; m + n \neq 0.
      \end{array}
      \right.
      \label{eq:cocycle2}
    \end{equation}
\end{proposition}

\Proof
This is straightforward using \myref{eq:note}. \qed

We shall denote by
$$
\hatDcal_q = \Dcal_q + \C C
$$
the central extension of $\Dcal_q\/$ corresponding to the cocycle
\myref{eq:cocycle2} so that the bracket of two elements from the
subspace $\Dcal_q\/$ is given by
$$
[ E_1, E_2 ] =
E_1 E_2 - E_2 E_1 +
\Psi ( E_1, E_2 ) C.
$$

\mysubsection
Let $\Dcal^a = \Dcal^a_1\/$, $\Dcal = \Dcal_1\/$, $\hatDcal =
\hatDcal_{1, \tr}\/$, $D = D_1\/$ ($= z \partial_z\/$).  As we
have seen $\Dcal^a\/$ is the associative algebra of all regular
differential operators on the punctured complex plane
$\C^\times\/$, i.e., operators of the form
\begin{equation}
  E = e_k ( z ) \partial^k_z +
  e_{k-1} ( z ) \partial^{k-1}_z + \cdots +
  e_0 ( z ), \qquad
  \mbox{where}\quad
  e_i ( z ) \in \C [z, z^{-1} ].
  \label{eq:puncplane}
\end{equation}

It is not difficult to see that the cocycle $\Psi\/$ given by
\myref{eq:cocycle2} is given by the following formula:
\begin{equation}
  \Psi ( f \partial^m_z, g \partial^n_z ) =
  \frac{m! n!}{( m + n + 1 )!} \Res_{z = 0} dz
  f^{(n+1)} ( z ) g^{(m)} ( z ),
  \label{eq:cocyle3}
\end{equation}
where as usual $f^{(n)}\/$ stands for $\partial^n_z f\/$.  This
cocycle appeared (probably for the first time) in \cite{KP}.
 It has been shown independently by several authors
(\cite{Li} and \cite{F} among them) that $\hatDcal\/$ is the unique, up to
isomorphism, non-trivial central extension of the Lie algebra
$\Dcal\/$ by a one-dimensional algebra.

It is, however, more convenient to write the differential
operators as linear combinations of elements of the form $z^k f (
D )\/$, where $f\/$ is a polynomial in $D\/$, since it is easier
to compute their product (cf.\ \myref{eq:multiplication}):
\begin{equation}
  ( z^m f ( D ) ) ( z^k g ( D ) ) =
  z^{m + k} f ( D + k ) g ( D ).
  \label{eq:compprod}
\end{equation}
The bracket in $\hatDcal\/$ is then given by
\begin{equation}
  [ z^r f ( D), z^s q ( D ) ] =
  z^{r + s} ( f ( D + s ) g ( D ) -
  f ( D ) g ( D + r ) +
  \Psi ( z^r f ( D ) , z^s g ( D ) ) ) C,
  \label{eq:bracket}
\end{equation}
where
\begin{equation}
  \Psi ( z^r f ( D ), z^s g ( D ) ) =
  \left\{
  \begin{array}{c@{\quad}l}
    \sum_{-r \leq j \leq -1} f ( j ) g ( j + r ) &
    \mbox{if}\; r = -s \geq 0\\
    0 & \mbox{if}\; r + s \neq 0.
  \end{array}
  \right.
  \label{eq:hatdcalbracket}
\end{equation}

\mysubsection
Consider now the associative algebra $\Dcalaq\/$, the
corresponding Lie algebra $\Dcal_q\/$ and its central extension
$\hatDcal_q\/$ in the case $q \neq 1\/$.  Introduce the following
basis of $\Dcalaq\/$:
$$
T_{m,n} = q^{\frac12 (m + 1)n}
( (q - 1) D_q + 1 )^n,\qquad
m \in \Z,\quad n \in \Z_+.
$$
Then we have
\begin{equation}
  T_{m,n} T_{m', n'} =
  q^{\frac12 ( m'n - mn' )} T_{m+m', n+n'}.
  \label{eq:basis}
\end{equation}
The cocycle \myref{eq:cocycle2} on the Lie algebra $\Dcal_q\/$
becomes:
\begin{equation}
  \Psi ( T_{m,n}, T_{m', n'} ) =
  \delta_{m,-m'} \frac{\sinh ( \hbar m (n + n'))}{\sinh (\hbar ( n+n'))},
  \label{eq:cocycle4}
\end{equation}
where we let $q = e^{\frac12 \hbar}\/$.  Consequently, the commutation
relations of the Lie algebra $\hatDcal_q\/$ become:
\begin{equation}
  [ T_{m,n}, T_{m', n'} ]=
  2 \sinh \left( \hbar ( m'n - mn' ) \right) T_{m + m', n + n'} +
  \delta_{m, -m'}
  \frac{\sinh ( \hbar m ( n + n' ) )}{\sinh ( \hbar ( n + n' ) )} C.
  \label{eq:commrel}
\end{equation}

\Remarks
(a) Commutation relations \myref{eq:commrel} correspond to the
automorphism $\sigma'_q\/$ of $\C [ w ]\/$ given by $\sigma'_q (
w ) = q w\/$ (which is equivalent to $\sigma_q\/$), and to the
trace being an evaluation map at $w = 1\/$.  The
evaluation map at $w = 0\/$ gives the cocycle
$\Psi_0 ( T_{m, n}, T_{m', n'} ) = m \delta_{m, -m'} \delta_{n,
-n'}\/$, which is equivalent to $\Psi\/$ due to the argument of
Remark 1.3(d).

(b) If we take $A = \C [ w, w^{-1} ]\/$, $\sigma ( x ) = q x\/$
where $q = e^{\frac12 \hbar} \neq 1\/$, and $\tr ( \sum a_i w^i )
= a_0\/$, then in the basis $T_{m,n} = q^{\frac12 mn} z^m w^n\/$
($m, n \in \Z\/$) we obtain the commutation relations of the
trigonometric Sin-Lie algebra:
$$
  [ T_{m,n}, T_{m',n'} ] =
  2 \sinh ( \hbar ( m' n - m n' ) ) T_{m + m', n + n'} +
  m \delta_{m, -m'} \delta_{n, -n'} C.
$$

\mysection{Lie algebras $\hatDcal\/$ and $\hatDcal^\Ocal\/$.}
\label{sec:lie-algebras}

\mysubsection
Let as before $D = z \partial_z\/$ and let
$$
L^n_k = z^k D^n \in \Dcal \subset \hatDcal\qquad
( k \in \Z,\; n \in \Z_+).
$$
Define the {\em order\/} and the {\em weight\/} by
$$
  \ord L^n_k = n, \quad w t L^n_k = k, \quad
  \ord C = w t C = 0.
$$
It is clear from \myref{eq:bracket} and \myref{eq:basis} that the
order defines a filtration of $\hatDcal\/$:
\begin{equation}
\hatDcal^0 \subset \hatDcal^1 \subset \hatDcal^2 \subset\cdots,
\label{eq:filtration}
\end{equation}
and the weight defines the principal $\Z\/$-gradation of
$\hatDcal\/$:
\begin{equation}
\hatDcal = \bigoplus_{j \in \Z} \hatDcal_j.
\label{eq:zgradationond}
\end{equation}
Note that we have:
\begin{eqnarray}
  \Psi ( L^0_r, L^0_s ) & = & \delta_{r, -s} r,\label{eq:grad1}\\
  \Psi ( L^1_r, L^1_s ) & = & -\delta_{r,-s} \frac{r^3 -
    r}{6},\label{eq:grad2}\\
  \Psi ( L^0_r, L^1_s ) & = & \Psi ( L^1_r, L^0_s ) =
    \delta_{r,-s} \frac{r(r-1)}{2}\quad if\; r \geq
    0.\label{eq:grad3}
\end{eqnarray}
It follows that $\hatDcal^0\/$ is isomorphic to the oscillator
Lie algebra:
\begin{equation}
  \brackets{L^0_r, L^0_s} = \delta_{r, -s} r C.
  \label{eq:oscillator}
\end{equation}
Furthermore, $\hatDcal^1\/$ contains a 1-parameter family of
Virasoro algebras $\Vir ( \beta )\/$, $\beta \in \C\/$,
(``complementary'' to $\hatDcal^0\/$) defined by
\begin{equation}
  L_k ( \beta ) = - \parens{L^1_k + \beta ( k + 1 ) L^0_k},
  \label{eq:virasoro1}
\end{equation}
so that
\begin{equation}
  \brackets{L_r (\beta), L_s (\beta)} =
  (r - s) L_{r+s} (\beta) + \delta_{r,-s}
  \frac{r^3-r}{12} C_{\Vir(\beta)}
  \label{eq:virasoro2}
\end{equation}
where
\begin{equation}
  C_{\Vir ( \beta )} = ( 12 \beta^2 - 2 ) C.
  \label{eq:cvir}
\end{equation}

\Remark
$z^{n+s} \partial^n_z = z^s  D ( D - 1 ) \cdots ( D - n + 1
)\/$.

\mysubsection
Let $\Ocal\/$ be the algebra of all holomorphic functions on
$\C\/$ with topology of uniform convergence on compact sets.  We
define a completion $\Dcal^{a \Ocal}\/$ of the
(associative) algebra of differential operators on $\C^\times\/$
by considering differential operators of infinite order of the
form $z^k f (D)\/$, where $f \in \Ocal\/$.  The usual product of
differential operators extends to $\Dcal^{a\Ocal}\/$:
\begin{equation}
  \left( z^r f(D) \right)
  \left( z^s g(D) \right) =
  z^{r+s} f(D + s) g (D),
  \label{eq:usual-product}
\end{equation}
where by $f (D + s)\/$ we mean the power series expansion in
$D\/$.  The principal gradation extends as well: $\Dcal^{a\Ocal}
= \bigoplus_{j \in \Z} \Dcal^{a\Ocal}_k\/$, where
$\Dcal^{a\Ocal}_k = \{ z^k f ( D ) | f ( w ) \in \Ocal \}\/$.
Identifying $\Dcal^{a\Ocal}_k\/$ with $\Ocal\/$ and
$\Dcal^{a\Ocal}\/$ with the direct sum of $\Dcal^{a \Ocal}_k\/$
as topological vector spaces, we make $\Dcal^{a\Ocal}\/$ a
topological associative algebra.  It is a completion of the
subalgebra $\Dcal^a\/$.

We denote by $\Dcal^\Ocal\/$ the corresponding (topological) Lie
algebra. Then the cocycle $\Psi\/$ extends by continuity from
$\Dcal\/$ to a 2-cocycle on $\Dcal^\Ocal\/$ by formula
\myref{eq:hatdcalbracket}.  We let
\begin{displaymath}
  \hatDcal^\Ocal = \Dcal^\Ocal \oplus \C C
\end{displaymath}
be the corresponding central extension.  Note that for elements
$z^r e^{\lambda D} ( r \in \Z, \lambda \in \C )\/$ the commutator
in $\hatDcal^\Ocal\/$ is especially simple:
\begin{equation}
  \left[
  z^r e^{\lambda D}, z^s e^{\mu D}
  \right] =
  \left(
  e^{\lambda s} - e^{\mu r}
  \right)
  z^{r+s} e^{(\lambda + \mu) D} +
  \delta_{r, -s}
  \frac{e^{-\lambda r} - e^{-\mu s}}{1 - e^{\lambda + \mu}} C.
  \label{eq:simple-commutator}
\end{equation}

\Remarks
(a) One may consider $z^k e^{\lambda D}\/$ as a generating series
for the $L^n_k\/$ and derive \myref{eq:bracket} and
\myref{eq:hatdcalbracket} by taking derivatives of
\myref{eq:simple-commutator}.

(b) Of course, $\Dcal^\Ocal\/$ is isomorphic to
$\tilde\Ocal_{\sigma_1}\/$, and $\hatDcal^\Ocal\/$ to
$\hat\Ocal_{\sigma_1, \tr}\/$.

(c) Consider the following traces on $\Ocal\/$:
$$
\begin{array}{rcl@{\quad}l}
  \tr_{a, b} f(w) & = & f(a) - f(b), & \mbox{where}\; a, b, \in \C,\\
  \tr^{[m]}_s f(w) & = & f^{(m)}(s), & \mbox{where}\; s \in \C,\;
  m \in \N
\end{array}
$$
Here and further $f^{(m)}\/$ stands for the $m\/$-th derivative
of $f(w)\/$.  We denote the corresponding
cocycles by  $\Psi_{a, b}:= \Psi_{\sigma_1, \tr_{a,b}}\/$ and
$\Psi^{(m)}_s := \Psi_{\sigma_1, \tr^{[m]}_s}\/$. On
$\Dcal^\Ocal\/$ these cocycles are nontrivial (in continuous
cohomology).  But, due to Remark 1.3(d) when restricted to
$\Dcal\/$ they become trivial.  Since
\begin{equation}
  ( \sigma_1 - 1 )
  \frac{e^{x w} - 1}{e^x - 1} =
  e^{x w}
\end{equation}
using Remark 1.3(d), we obtain the following explicit formulas
for these trivial cocycles on $\Dcal\/$:
\begin{eqnarray}
  \Psi_{a,b}( z^k f(D), z^r g(D) ) & = &
    \delta_{k, -r} \Lambda_{a, b} ( [ z^k f(D), z^{-k} g(D) ] ),
    \label{eq:trivcocycles1}\\
  \Psi^{(m)}_s ( z^k f(D), z^r g(D) ) & = &
    \delta_{k, -r} \Lambda^{(m)}_s ( [ z^k f(D), z^{-k} g(D) ] ),
    \label{eq:trivcocycles2}
\end{eqnarray}
where $\Lambda_{a,b}\/$ and $\Lambda^{(m)}_s\/$ are the
linear functions on $\C[w]\/$ defined by the following generating
series in $x\/$:
\begin{equation}
  \Lambda_{a,b} ( e^{x w} ) =
  \frac{e^{a x} - e^{b x}}{e^x - 1},\qquad
  \Lambda^{(m)}_s ( e^{x w} ) =
  \frac{x^m e^{s x}}{e^x - 1}.
  \label{eq:generating}
\end{equation}

\mysubsection
The following theorem describes closed ideals of $\Dcal^\Ocal\/$
($\resp \hatDcal^\Ocal\/$).

\begin{theorem}
  (a) The center $Z$ of $\Dcal^\Ocal$ consists of elements of
  the form $f(D)$, where $f(w) \in \Ocal$ is a 1-periodic
  function (i.e., $f(w+1) = f(w)$).  The center of
  $\hatDcal^\Ocal$ is $\hat Z = Z \oplus \C C$.

  (b) Let $I$ be an ideal of $\Ocal$ which is invariant under
  the translation $w \mapsto w + 1$, and let $I' = \{ f(D) -
  f(D+1)| f(w) \in I \}$ ($\resp \hat I' = \{ f(D) - f(D+1) +
  f(0) C | f(w) \in I \}$).  Let $Y$ be a subspace of $Z$
  ($\resp \hat Z$).  Let $I^{(k)} = z^k I \subset
  \Dcal^\Ocal$ for $k \neq 0\/$ and let $I^{(0)} = I' + Y$
  ($\resp = \hat I' + Y$).  Then
  $$
  J ( I, Y) = \bigoplus_{k \in \Z} I^{(k)}
  $$
  is a closed ideal of $\Dcal^\Ocal$ (resp.\ $\hatDcal^\Ocal$).

  (c) Every closed ideal of $\Dcal^\Ocal$ (resp.\ of
  $\hatDcal^\Ocal$) is one of the $J (I, Y)$.
\end{theorem}

\Proof
The statements (a) and (b) are clear.  We shall prove (c) for
$\Dcal^\Ocal\/$, the proof for $\hatDcal^\Ocal\/$ being the same.
Let $J\/$ be a closed ideal of the Lie algebra $\Dcal^\Ocal\/$.
Since $J\/$ is $\ad D\/$-stable, it follows that $J\/$ is a
graded ideal:
$$
J = \bigoplus_{k \in \Z} z^k I_k,
$$
where $I_k\/$ is a closed subspace of $\Ocal\/$.  We have
$$
\left[
D^2, z^k f(D)
\right] =
2k z^k D f(D) + k^2 z^k f(D).
$$
It follows that $w I_k \subset I_k\/$ if $k \neq 0\/$, i.e., that
$I_k\/$ for $k \neq 0\/$ is an ideal of $\Ocal\/$.  Furthermore,
we let for $k \in \Z\/$ and an ideal $I\/$ of $\Ocal\/$:
$$
I [ k ] =
\left\{
f \in \Ocal \mathop{|} f (w + k) \in I
\right\}.
$$
We claim that
$$
  I_k [ \pm 1] + I_k \subset
  I_{k \pm 1}\quad \mbox{if}\quad
  k \neq 0\; \mbox{and}\; k \pm 1 \neq 0.
  \eqno(2.3.1)_\pm
$$
\setcounter{equation}{1}%\label{eq:idealIO}
Indeed, since
$$
\left[
z^{\pm 1}, z^k f(D)
\right] =
z^{k \pm 1}
\left(
f(D) - f(D \pm 1)
\right),
$$
we see that if $f(w) \in I_k\/$ then $f(w) - f(w \pm 1) \in I_{k
\pm 1}\/$.  Since $I_k\/$ and $I_{k\pm 1}\/$ are ideals, $w f(w)
\in I_k\/$ and $w f(w) - (w \pm 1) f (w \pm 1) \in I_{k \pm
1}\/$.  Thus, $f(w), f(w \pm 1) \in I_{k \pm 1}\/$, completing
the proof of $(2.3.1)_\pm$.  We conclude, in particular,
that
\begin{equation}
  I_1 = I_2 = \ldots\quad \mbox{and}\quad
  I_{-1}= I_{-2} = \ldots\ .
  \label{eq:conclude}
\end{equation}

Next, we prove that
\begin{equation}
  I_k = I_k [n]\; \mbox{for all}\; n \in \Z, \; \mbox{provided
  that}\; k \neq 0.
  \label{eq:Ikn-in-Z}
\end{equation}
Indeed, due to \myref{eq:conclude} we may assume that $|k| \geq
2\/$, so that both numbers $k + 1\/$ or $k - 1\/$ are non-zero.
Applying $(2.3.1)_+\/$ to $I_k\/$ and $(2.3.1)_-\/$ to
$I_{k-1}\/$, we get
$$
I_k [1] + I_k \subset I_{k+1}\quad
\mbox{and}\quad
I_{k+1} [-1] + I_{k+1} \subset I_k.
$$
It follows that
$
  I_k = \left( I_k [1] \right) [-1] \subset I_{k+1} [-1] \subset
  I_k,
\/$ and
$
  I_k = \left( I_k[-1] \right) [1] \subset I_k [1] \subset
  I_{k+1} = I_k.
\/$
Hence $I_k = I_k [ \pm 1 ]\/$ proving \myref{eq:Ikn-in-Z}, which
means that each $I_k\/$ is invariant under the translation $w
\mapsto w + 1\/$.

In order to complete the proof of the theorem, it remains to show
that
\begin{eqnarray}
  I_1 & = & I_{-1} \label{eq:show1}\\
  I_0 & \supset & I' \label{eq:show2}\\
  I_0 & \subset & I' + Z. \label{eq:show3}
\end{eqnarray}
First, we prove that $I_{-1} \subset I_1\/$; the reverse
inclusion is proved similarly.  Let $f(w) \in I_{-1}\/$; we have:
$$
\left[
 z, [z, z^{-1} f(D)]
\right] =
z \left(
   f(D) - 2f(D+1) + f(D+2)
  \right).
$$
Hence $f(w) - 2f(w+1) + f(w+2) \in I_1\/$.  Considering $w f(w)
\in I_{-1}\/$, we conclude that $f(w+2) - f(w+1) \in I_1\/$ and
$f(w+1) - f(w) \in I_1\/$.  Considering $w f(w)\/$ once more, we
conclude that $f(w) \in I_1\/$, proving \myref{eq:show1}.

The inclusion \myref{eq:show2} follows from the inclusions $[z,
I_{-1}] \subset I_0\/$ and $[z^{-1}, I_1] \subset I_0\/$.
Finally, in order to prove \myref{eq:show3}, note that the map
$\varphi := ( \ad z)^2 : z^{-1}I_{-1} \rightarrow z I_1\/$ is
surjective (this follows from the proof of \myref{eq:show1}).  Let
now $f \in I_0\/$ and let $g \in z^{-1} I_{-1}\/$ be a pre-image
of $[ z, f(D) ]\/$ under the map $\varphi\/$.  Then $[z, g] \in
I'\/$ and $[ f - [z, g], z] = 0\/$, hence $f - [z, g] \in Z\/$,
proving \myref{eq:show3}.\qed

We have the following corollary of the proof:

\begin{corollary}
  The Lie algebra $\Dcal / \C$ is simple.
\end{corollary}

\mysubsection
Consider a parabolic subalgebra $\frak p\/$ of $\hatDcal\/$:
$$
\pfrak = \bigoplus_{j \in \Z} \frak p_j,\; \mbox{where}\;
\frak p_j = \hatDcal_j\; \mbox{for}\; j \geq 0\; \mbox{and}\;
\frak p_j \neq 0\; \mbox{for some}\; j < 0.
$$

For each positive integer $k\/$ we have:  $\pfrak_{-k} = z^{-k}
I_{-k}\/$, where $I_{-k}\/$ is a subspace of $A = \C [ w ]\/$.
Since
$$
\left[
  f ( D ), z^{-k} P ( D )
\right] =
z^{-k}
\left(
  f ( D - k ) - f ( D )
\right)
P ( D ),
$$
we see that $I_{-k}\/$ is an ideal of the polynomial algebra
$A\/$.  It is clear that $I_{-k} \neq 0\/$ for all $k = 1, 2,
\ldots\/$.  Let $b_{k} ( w )\/$ be the monic (i.e., with the
leading coefficient equal to 1) polynomial which is a generator
of the ideal $I_{-k}\/$.  Thus, to a parabolic subalgebra
$\pfrak\/$ we have associated a sequence of monic polynomials
$b_1 = b_1 ( w )\/$, $b_2 = b_2 ( w )\/$, \ldots.  The polynomial
$b_k\/$, $k = 1, 2, \ldots\/$, are called the {\em characteristic
polynomials of $\pfrak\/$\/}.

\begin{lemma}
  Let $\{ b_k \}_{k \in \N}$ be the sequence of characteristic
  polynomials of a parabolic subalgebra $\pfrak$ of the Lie
  algebra $\hatDcal$.  Then
  \begin{enumerate}
  \item[(a)] $b_k ( w )$ divides $b_{k+1} ( w )$ and $b_{k+1}
    ( w + 1 )$ for all $k \in \N$.

  \item[(b)] $b_{k+l} ( w )$ divides $b_k ( w - l) b_l ( w )$
    for all $k, l \in \N$.
  \end{enumerate}
\end{lemma}

\Proof
Since $[ z , z^{-k-1} b_{k+1} ( D ) ] = z^{-k} ( b_{k+1} ( D ) -
b_{k+1} ( D + 1) )\/$, we see that $b_k ( w )\/$ divides $b_{k+1} (
w ) - b_{k+1} ( w + 1 )\/$.  Since $[ z, z^{-k - 1} D b_{k+1} ( D
) ] = z^{-k} ( D b_{k + 1} ( D ) - ( D + 1 ) b_{k + 1} ( D + 1 )
)\/$, we see that $b_k ( w )\/$ divides $w ( b_{k + 1} ( w ) -
b_{k+1} ( w + 1 ) ) + b_{k + 1} ( w + 1 )\/$, proving (a).

The proof of (b) is similar by computing the commutators
$[ z^{-k} b_k ( D ), z^{-l} b_l ( D ) ]\/$ and $[ z^{-k} b_k ( D
),$ $z^{-l} D b_l ( D ) ]\/$.\qed

Given a monic polynomial $b(w)\/$, we let
$$
b^{\rm min}_k (w) =
b(w) b(w-1)\ldots b(w - k + 1),
$$
$$
b^{\rm max}_k ( w ) =
lcm \left\{ b(w), b(w - 1), \ldots, b(w - k + 1) \right\}.
$$
It is
easy to see that there exist (unique) parabolic subalgebras,
which we denote by $\pfrak_{\rm min} (b)\/$ and $\pfrak_{\rm max} (b)\/$,
for which the characteristic polynomials are $\{ b^{\rm min}_k
(w) \}\/$ and $\{ b^{\rm max}_k ( w )\}\/$ respectively.  We
clearly have
\begin{equation}
  \dim \hatDcal_{-k} / \pfrak_{\rm min} (b)_{-k} = k \deg b.
  \label{eq:charpoly}
\end{equation}

Lemma 2.4 implies the following.

\begin{proposition}
  Let $b$ be a monic polynomial and let $\pfrak$ be a
  parabolic subalgebra such that $b_1(w) = b$.  Then
  $$
  \pfrak_{\rm min} ( b ) \subset \pfrak \subset \pfrak_{\rm max}
  ( b ).
  $$
  In particular, if difference of any two distinct roots of $b$
  is not an integer, then
  $$
  \pfrak = \pfrak_{\rm min} ( b ) = \pfrak_{\rm max} ( b ).\qed
  $$
\end{proposition}

\Remark
$\pfrak_{\rm min} ( b ) = \pfrak( (b) )\/$ (cf.\ \myref{eq:parsub}).

\mysubsection
Given a monic polynomial $b = b(w)\/$, consider the following
subspace of $\hatDcal_0\/$:
$$
\hatDcal^b_0 =
\left\{
b ( D ) g ( D ) - b ( D + 1 ) g ( D + 1 ) +
b ( 0 ) g ( 0 ) C\, |\, g ( w ) \in \C[w]
\right\}.
$$

In order to study modules over $\hatDcal\/$ induced from its
parabolic subalgebras, we need the following proposition.

\begin{proposition}
  Let $\pfrak$ be a parabolic subalgebra of $\hatDcal$ and
  let $b = b(w)$ be its first characteristic polynomial.  Then
  $$
  \left[
  \pfrak, \pfrak
  \right] =
  \left(
  \bigoplus_{k \neq 0} \pfrak_{-k}
  \right) \bigoplus \hatDcal^b_0.
  $$
  In particular,
  \begin{equation}
    \dim \pfrak / [ \pfrak, \pfrak ] =
    \dim \pfrak_0 / [\pfrak, \pfrak]_0 = \deg b(w).
    \label{eq:polychar}
  \end{equation}
\end{proposition}

\Proof
Note that $[ \pfrak, \pfrak ]_0 = [\pfrak_1, \pfrak_{-1} ]\/$ and
that $[z f ( D + 1 ), z^{-1} b ( D ) g ( D ) ] = b ( D ) f ( D )
g ( D ) -$ \linebreak $b ( D + 1) f ( D + 1 ) g ( D + 1 ) + b ( 0
) f ( 0 ) g ( 0 ) C\/$.  The rest is straightforward.\qed

\mysubsection
Let $\gfrak\/$ be a finite-dimensional semisimple Lie algebra
over $\C\/$ and let $\gfrak = \bigoplus_\alpha \gfrak_\alpha\/$
be its root space decomposition with respect to a Cartan
subalgebra $\gfrak_0\/$.  An embedding $\gfrak \subset
\Dcal^\Ocal\/$ is called {\em graded\/} if $\gfrak_\alpha \subset
\Dcal^\Ocal_{k ( \alpha )}\/$ for all $\alpha\/$.

\begin{proposition}
  (a) The  graded embeddings in $\Dcal^\Ocal$ of the Lie
  algebra $sl_2 ( \C )$ with the standard basis $E$, $H$,
  $F$ are parametrized by $k \in \Z \backslash \{ 0 \}$
  and  by $k$-periodic functions $f,g \in \Ocal$ as follows:
  $$
  H = \frac 2k D + f ( D ),\quad
  E = z^k x ( D ),\quad
  F = z^{-k} y ( D ),
  $$
  where
  $$
  x ( D ) y ( D + k ) =
  - \left( \frac Dk \right)^2 + \frac Dk \Bigl( f ( D ) + 1 \Bigr)+
  g ( D ).
  $$

  (b) The only graded embeddings of $sl_2 ( \C )$ in $\Dcal$
  are as follows ($k \in \Z \backslash \{ 0 \}$ , $ \lambda, \mu,
  \in \C$):
  $$
  H = \frac 2k D + \lambda,\quad
  E = z^k x ( D ),\quad
  F = z^{-k} y ( D ),
  $$
  with the following four possibilities for $x ( D )$ and $y (
  D )$, where $\mu \in \C$:
  \begin{enumerate}
    \renewcommand{\labelenumi}{(\roman{enumi})}
  \item $x ( D ) = \frac 1k D^2 + \frac 1k ( \lambda + 1 ) D +
    \mu$, $y ( D ) = - \frac 1k$;

  \item $x ( D ) = \frac 1k D - \mu$, $y ( D ) = - \frac 1k D -
    \lambda - \mu - 1$;

  \item[\em (iii) and (iv)] are  obtained from (i) and (ii) by
    the substitution $x' ( D ) = y ( D - k )$, $y' ( D ) = x ( D
    + k )$.
  \end{enumerate}

  (c) A semisimple Lie algebra of rank $\geq 2\/$ has no graded
  embeddings in $\Dcal^\Ocal\/$.
\end{proposition}

\Proof
Note that the equation
$$
\Bigl[ h ( D ), \, z^k x ( D ) \Bigr] = \lambda z^k x ( D )
$$
implies that
$$
h ( D + k ) - h ( D ) = \lambda.
$$
All solutions of the latter equation are $h ( 0 ) = \frac \lambda
k D + f ( D )\/$, where $f ( D + k ) = f ( D )\/$.  Now (a)
easily follows. (b) follows from (a).  If rank $\frak g \geq 2\/$, we always
can find
an element $h \in \frak g_0\/$ such that $\alpha ( h ) / \beta (
h )\/$ is an irrational number for two distinct roots $\alpha\/$
and $\beta\/$.  Hence (a) implies (c).\qed

\Remarks
(a)  Let $L_n\/$ denote the subalgebra of operators of
$\Dcal^\Ocal\/$ leaving invariant the subspace $\sum^n_{k=0} \C z
^k\/$ of $\C [ z, z^{-1} ]\/$, and let $I_n\/$ denote the ideal
of $L_n\/$ of operators acting on this subspace trivially.  By Jacobson's
density theorem we have an exact sequence of associative
algebras:
$$
0 \rightarrow I_n \rightarrow L_n \rightarrow \mbox{Mat}_n ( \C )
\rightarrow 0.
$$
Proposition 2.6(b) shows that this is a non-split exact sequence.

(b) It follows from the proof of the proposition that $\ad
\Dcal^\Ocal_0\/$ is not diagonalizable on $\Dcal^\Ocal\/$.

\mysection{Interplay between $\hatDcal\/$ and
$\hatgl ( \infty ) [ m ]\/$}
\label{sec:interplay}

\mysubsection
Let $R\/$ be an associative algebra over $\C\/$.  Denote by
$R^\infty\/$ with a fixed basis $\{ v_j \}_{j \in \Z}\/$.  As
usual, define the operators $E_{ij}\/$ by
\begin{equation}
  E_{ij} v_k = \delta_{jk} v_i.
  \label{eq:assocalg}
\end{equation}
Denote by $\widetilde M (\infty , R)\/$ the
associative subalgebra of $\End R^\infty\/$ consisting of all
operators $\sum_{i,j \in \Z} a_{ij} E_{ij}\/$ whose matrices
$( a_{ij} )_{i, j \in \Z}\/$ have a finite number of non-zero
diagonals.  Letting $\deg E_{ij} = j -
i\/$ defines the principal $\Z\/$-gradation:
\begin{equation}
  \Mtilde ( \infty , R ) = \bigoplus_{j \in \Z} \Mtilde ( \infty
  , R )_j.
  \label{eq:gradation}
\end{equation}

Fix $s \in \C\/$ and a nilpotent element $t \in R\/$.  Consider
the free $R\/$-module $R [ z, z^{-1} ] z^s\/$ and identify it
with $R^\infty\/$ by choosing the basis $v_j = z^{-j+s}\/$, $j
\in \Z\/$.  By associating to an element $z^k f(D) \in \Dcal^a\/$
the operator $z^k f(D + t)\/$ on $R [ z, z^{-1} ] z^s\/$, we
obtain an embedding $\varphi_{s, t} : \Dcal^a \hookrightarrow
\Mtilde ( \infty , R )\/$ of associative algebras over $\C\/$,
which is compatible with the principal gradations.  Explicitly:
\begin{equation}
  \varphi_{s, t}
  \left( z^k f(D) \right) =
  \sum_{j \in \Z} f ( -j + s + t ) E_{j -k, j}.
  \label{eq:comp-grads}
\end{equation}

The homomorphism $\varphi_{s, t} : \Dcal^a \rightarrow \widetilde M (
\infty , R )\/$ extends via \myref{eq:comp-grads} to a homomorphism
\begin{displaymath}
  \varphi_{s, t} : \Dcal^{a \Ocal} \rightarrow \widetilde M (
  \infty , R ).
\end{displaymath}

\mysubsection
Given a non-negative integer $m\/$, consider the algebra of
truncated polynomials $R_m = \C [ t ] / ( t^{m + 1} )\/$, and let
$\Mtilde ( \infty ) [ m ] = \Mtilde ( \infty, R_m )\/$.  We
denote the homomorphism $\varphi_{s, t} : \Dcal^{a \Ocal}
\rightarrow \Mtilde ( \infty ) [ m ]\/$ given by
\myref{eq:comp-grads} by $\varphi^{[m]}_s\/$.  By Taylor's
formula we have:
\begin{equation}
  \varphi^{[m]}_s \left( z^k f(D) \right) =
  \sum^m_{i = 0} \sum_{j \in \Z}
  \frac{f^{(i)} ( -j + s )}{i !}
  t^i E_{j-k, j}.
  \label{eq:taylor1}
\end{equation}
Let
$$
I^{[m]}_s =
\left\{
f \in \Ocal | f^{(i)} ( n + s ) = 0\quad \mbox{for all}\quad
n \in \Z\quad \mbox{and all}\quad i = 0, \ldots, m
\right\},
$$
and let $J^{[m]}_s = \bigoplus_{k \in \Z} z^k I^{[m]}_s \in
\Dcal^{a \Ocal}\/$.  We clearly have:
\begin{equation}
  \Ker \varphi^{[m]}_s = J^{[m]}_s.
  \label{eq:taylor2}
\end{equation}

Fix now $\vec{s} = (s_1, \ldots, s_N ) \in \C^N\/$ such that $s_i
- s_j \not\in \Z\/$ if $i \neq j\/$, and fix $\vec{m} = ( m_1,
\ldots, m_N ) \in \Z^N_+\/$.  Let $\Mtilde ( \infty ) [ \vec m ]
= \bigoplus^N_{i = 1} \Mtilde ( \infty ) [ m_i ]\/$.  Consider
the homomorphism
$$
\varphi^{[\vec{m}]}_{\vec{s}} = \bigoplus_i
\varphi^{[m_i]}_{s_i} :
\Dcal^{a \Ocal} \rightarrow \Mtilde ( \infty ) [ \vec m ].
$$

\begin{proposition}
  We have an exact sequence of $\Z$-graded associative
  algebras:
  $$
  0 \rightarrow J^{[\vec{m}]}_{\vec{s}} \rightarrow \Dcal^{a
  \Ocal} \stackrel{\varphi^{[\vec{m}]}_{\vec{s}}}{\longrightarrow}
  \Mtilde ( \infty ) [ \vec m ] \rightarrow 0,
  $$
  where $J^{[\vec m]}_{\vec s} = \bigcap^N_{i = 1} J^{[m_i]}_{s_i}$.
\end{proposition}

\Proof
It is clear from \myref{eq:taylor2} that $\Ker \varphi^{[\vec
m]}_{\vec s} = J^{[ \vec m]}_{\vec s}\/$.  The surjectivity of
$\varphi^{[\vec m]}_{\vec s}\/$ follows from the following
well-known fact:  for every discrete sequence of points in $\C\/$
and a non-negative integer $m\/$ there exists $f(w) \in \Ocal\/$
having prescribed values of its first $m\/$ derivatives at these points.

\mysubsection
We denote by $\gltilde ( \infty ) [ m ]\/$ the $\Z\/$-graded Lie
algebra over $\C\/$ corresponding to the associative algebra
$\Mtilde ( \infty ) [ m ]\/$ viewed as an algebra over $\C\/$.
Consider the following 2-cocycle on $\gltilde ( \infty ) [ m ]\/$
with values in $R_m\/$:
\begin{equation}
  C ( A, B ) = \tr [ J, A ] B
  \label{eq:uniquecohom}
\end{equation}
where $J = \sum_{i \leq 0} E_{ii}\/$, and
denote by $\hatgl ( \infty ) [ m ]= \gltilde ( \infty ) [ m ] +
R_m\/$ the corresponding central extension.  The $\Z\/$-gradation
of this Lie algebra extends from $\gltilde ( \infty ) [ m ]\/$ by
letting $wt R_m = 0\/$.

The homomorphism $\varphi_s^{[ m ]}\/$ of the associative algebras defines a
homomorphism of the corresponding Lie algebras, which we denote
by the same letter:
\begin{displaymath}
  \varphi_s^{[ m ]} : \Dcal \rightarrow \gltilde ( \infty ) [ m ] \quad
  \mbox{and}\quad
  \varphi_s^{[ m ]}: \Dcal^\Ocal \rightarrow \gltilde ( \infty ) [ m ].
\end{displaymath}

Denote by $\Psi^{[ m ]}_s\/$ the restriction of the cocycle $C\/$
given by \myref{eq:polychar} to $\varphi^{[m]}_s ( \Dcal^\Ocal
)\/$.  This gives us the following $R_m\/$-valued cocycle on
$\Dcal^\Ocal\/$:
\begin{equation}
  \Psi^{[m]}_s = \Psi + \Psi_{s, 0} +
  \sum^m_{j = 1} \Psi^{(j)}_s
  \frac{t^j}{j!},
  \label{eq:Rm-valued}
\end{equation}
where the cocycle $\Psi\/$ is given by \myref{eq:hatdcalbracket} and
the cocycles $\Psi_{s, 0}\/$ and $\Psi^{(j)}_s\/$ are
defined in Remark 2.2(c).  Using (\ref{eq:trivcocycles1}--6), we
thus obtain the following proposition.

\begin{proposition}
  The $\C$-linear map $\hat\varphi^{[m]}_s : \hatDcal
  \rightarrow \hatgl ( \infty ) [ m ]$ defined by
  \begin{equation}
    \hat\varphi^{[m]}_s
    \left|_{\hatDcal_j}
    = \varphi^{[m]}_s
    \right|_{\Dcal_j}\quad \mbox{if}\quad
    j \neq 0,
  \label{eq:linearmap1}
  \end{equation}
  \begin{equation}
    \hat\varphi^{[m]}_s ( e^{x D} ) =
    \varphi^{[m]}_s ( e^{x D} ) -
    \frac{e^{s x} - 1}{e^x - 1} -
    \sum^m_{j = 1} \frac{x^j e^{s x}}{e^x - 1} t^j / j!,\quad
    \hat\varphi^{[m]}_s ( C ) = 1 \in R_m
  \label{eq:linearmap2}
  \end{equation}
is a homomorphism of Lie algebras over $\C\/$.
\end{proposition}

\mysubsection
Define an automorphism $\nu\/$ of the algebra $\widetilde M (
\infty, \C )\/$ by letting
\begin{displaymath}
  \nu ( E_{ij} ) = E_{i + 1, j + 1}.
\end{displaymath}
Let $\varphi_s = \varphi_{s, 0} : \Dcal^{a \Ocal} \rightarrow
\Mtilde ( \infty, \C )\/$ (see Section 3.1).
Then we have
\begin{equation}
  \varphi_{s + 1} \parens{z^k f ( D )} = \nu \varphi_s
  \parens{z^k f ( D )}.
  \label{eq:autoalg}
\end{equation}

\Definition
A {\em monodromic loop\/} is a map $f : \C \rightarrow \widetilde M (
\infty, \C )\/$ such that
\begin{enumerate}
\renewcommand{\labelenumi}{(\roman{enumi})}
\item $f\/$ is holomorphic on $\C\/$, i.e., $f ( w ) = \sum_{ij}
  f_{ij} ( w ) E_{ij}\/$, where $f_{ij} \in \Ocal\/$,

\item $f ( w + 1 ) = \nu f ( w )\/$, i.e., $f_{ij} ( w + 1 ) =
  f_{i + 1, j + 1} ( w )\/$.
\end{enumerate}
We let $\Lcal_\nu \widetilde M ( \infty )\/$ denote the
associative algebra of all monodromic loops.  It clearly inherits
from $\widetilde M ( \infty, \C )\/$ the principal gradation.

Define a linear map $\varphi : \Dcal^{a \Ocal} \rightarrow
\Lcal_\nu \widetilde M ( \infty )\/$ by letting
\begin{equation}
  \varphi ( E ) = \mbox{loop}\; \braces{s \mapsto \varphi_s ( E )}.
  \label{eq:linearmap}
\end{equation}
This is a homomorphism of associative algebras.  The converse
homomorphism $\varphi^{-1}\/$ is constructed as follows.  Given
$f ( w ) = \sum_j f_j ( w ) E_{j, j+k} \in \Lcal_\nu \widetilde M
( \infty )\/$ (a monodromic loop concentrated on $k\/$-th
diagonal), we let
\begin{equation}
  \varphi^{-1} \parens{f ( w )} = z^k f_0 ( D ).
  \label{eq:convhomo}
\end{equation}
Thus we obtain the following result.

\begin{proposition}
  The map $\varphi$ is an isomorphism of $\Z$-graded
  associative algebras
  \begin{displaymath}
    \varphi : \Dcal^{a \Ocal} \simrightarrow \Lcal_\nu \widetilde M
    ( \infty )
  \end{displaymath}
\end{proposition}

\Remark
Monodromic loops are sections of the vector bundle on the
cylinder $\C / \Z\/$ with fiber $\widetilde M ( \infty, \C )\/$
and transition function $\nu\/$ in a small neighbourhood of the
line $\re w = 1\/$.

Denote by $\Lcal_\nu \gltilde ( \infty)\/$ the Lie algebra
obtained from $\Lcal_\nu \Mtilde ( \infty )\/$ by taking the
usual bracket.  For each $s \in \C\/$ define a 2-cocycle $C_s\/$
on this Lie algebra by
\begin{equation}
  C_s \left( f(w), g(w) \right) = C
  \left(
    f ( s ), g ( s )
  \right),
  \quad\mbox{where}\quad
  f(w), g(w) \in \Lcal_\nu \gltilde ( \infty ).
  \label{eq:cocyleonlie}
\end{equation}
It is easy to see that
under the isomorphism
$$
\varphi : \Dcal^\Ocal \simrightarrow \Lcal_\nu \gltilde ( \infty )
$$
given by Proposition 2.4, the cocycle $C_s\/$ induces the
following cocycle on $\Dcal^\Ocal\/$:
$$
\Psi_s
\left(
  z^k f ( D ), z^m g ( D )
\right) =
\left\{
\displaystyle
\begin{array}{cl}
  \displaystyle
  \sum_{-k \leq j \leq -1}  f ( j + s ) g ( j + k + s) &
    \mbox{if}\; k = -m \geq 0,\\
  0 &
    \mbox{if}\; k + m \neq 0.
\end{array}
\right.
$$
Denote by $\Lcal_\nu \gltilde ( \infty )^\wedge\/$ the central
extension of $\Lcal_\nu \gltilde ( \infty )\/$ corresponding to
the cocycle $C_0\/$.  Then the isomorphism $\varphi : \Dcal^\Ocal
\simrightarrow \Lcal_\nu \gltilde ( \infty )\/$ lifts to the
isomorphism $\hat\varphi : \hatDcal^\Ocal \simrightarrow
\Lcal_\nu \gltilde ( \infty )^\wedge\/$.

\mysection{Quasifinite highest weight modules over $\hatDcal\/$.}

\mysubsection
Let $b\/$ be a monic polynomial and let $\lambda \in
\hatDcal^\ast_0\/$ be such that $\lambda |_{\hatDcal^b_0} = 0\/$
(see Section 2.5).  Consider the parabolic subalgebra $\pfrak\/$
whose first characteristic polynomial is $\bfrak\/$ and denote by
$ = \pfrak_{\rm max} ( b )\/$ and denote by $M ( \lambda ; b )\/$
the generalized Verma module $M ( \hatDcal, \pfrak, \lambda )\/$.

\Definition
A Verma module $M ( \lambda )\/$ over $\hatDcal\/$ is called
{\em highly degenerate\/} if there exists a singular vector in $M
( \lambda )_{-1}\/$.

The following proposition follows from Propositions 2.4 and 2.5
and formula \myref{eq:charpoly}.

\begin{proposition}
  The following conditions on $\lambda \in \hatDcal^\ast_0$ are
  equivalent:
  \begin{enumerate}
  \item[(i)] $M ( \lambda )$ is highly degenerate;
  \item[(ii)] $L(\lambda)$ is quasifinite;
  \item[(iii)] $L ( \lambda )$ is a quotient of a
    generalized Verma module $M ( \lambda; b)$ for some monic
    polynomial $b$.\qed
  \end{enumerate}
\end{proposition}

Let $L ( \lambda )\/$ be a quasifinite irreducible highest weight
module over $\hatDcal\/$.  According to Proposition 4.1, $( z
^{-1} b ( D ) ) v_\lambda = 0\/$ for some monic polynomial $b ( w
)\/$.  Such monic polynomial of minimal degree is called the {\em
characteristic polynomial\/} of $L ( \lambda )\/$.  Note that $L
( \lambda )$ is a quotient of $M ( \lambda; b)$, where $b$
is the characteristic polynomial of $L ( \lambda )$.

\mysubsection
We shall characterize $\lambda \in \hatDcal^\ast_0\/$ by its labels
$\Delta_n = - \lambda ( D^n )\/$ and the central charge $c =
\lambda ( C )\/$.  Introduce the generating series
$$
\Delta_\lambda ( x ) =
\sum^\infty_{n = 0} \frac{x^n}{n!} \Delta_n.
$$

Recall that a {\em quasipolynomial\/} is a linear combination of
functions of the form $p ( x ) e^{\alpha x}\/$, where $p(x)\/$ is
a polynomial and $\alpha \in \C\/$.  Recall the following
well-known fact.

\begin{lemma}
  A formal power series is a quasipolynomial if and only if it
  satisfies a non-trivial linear differential equation with
  constant coefficients.
\end{lemma}

\begin{theorem}
  A $\hatDcal$-module $L ( \lambda )$ is quasifinite if and only
  if
  $$
  \Delta_\lambda ( x ) = \frac{\phi ( x )}{e^x - 1},
  $$
  where $\phi(x)$ is a quasipolynomial such that $\phi( 0 ) = 0$.
\end{theorem}

\Proof
It follows from Propositions 4.1 and 2.5 that $L ( \lambda )\/$
is quasifinite if and only if there exists a monic polynomial
$$
b ( w ) = w^N + f_{N-1} w^{N-1} + \cdots + f_0
$$
such that for all $s = 0, 1, \ldots\/$ we have:
$$
\lambda
\left(
D^s b(D) - (D+1)^s b(D+1) + b(0) \delta_{s,0} C
\right) = 0.
$$
This condition can be rewritten as follows:
\begin{equation}
  \sum^N_{n=0} f_n F_{n+s} = 0\; \mbox{for all}\; s = 0, 1, \ldots,
  \label{eq:monic-poly1}
\end{equation}
where
\begin{equation}
  F_n = \delta_{n,0} c +
  \sum^{n-1}_{j=0} \left( {n \atop j} \right) \Delta_j.
  \label{eq:monic-poly2}
\end{equation}
Introducing the generating series
$$
F ( x ) = \sum^\infty_{n = 0} \frac{x^n}{n!} F_n,
$$
we may rewrite \myref{eq:monic-poly1} in the form
\begin{equation}
  \left(
  \sum^N_{n=0} f_n \left(
                   \frac{d}{dx}
                   \right)^n
  \right) F ( x ) = 0.
  \label{eq:generating1}
\end{equation}
Thus, by Lemma 4.2, $L ( \lambda )\/$ is quasifinite if and only
if $F(x)\/$ is a quasipolynomial.

But \myref{eq:monic-poly2} can be rewritten in terms of
generating series as follows:
\begin{equation}
  c - F ( x ) =
  \left(
    e^x - 1
  \right) \Delta_\lambda ( x ).
  \label{eq:generating2}
\end{equation}
The theorem follows.\qed

{}From the proof of the theorem we obtain the following

\begin{corollary}
  Let $L ( \lambda )$ be a quasifinite irreducible highest weight
  module over $\hatDcal\/$, and let $b ( w )$ be its characteristic
  polynomial.  By Theorem 4.2, $F ( x ) = ( 1 - e^x )
  \Delta_\lambda ( x ) + c$ is a quasipolynomial.  Let $F^{(N)} +
  f_{N-1} F^{(N-1)} + \cdots + f_0 = 0$ be the minimal order linear
  differential equation with constant coefficients satisfied by $F
  ( x )$.  Then $b ( w ) = w^N + f_{N-1} w^{N-1} + \cdots +
  f_0$.\qed
\end{corollary}

\mysubsection
In this section we show that any quasifinite $\hat\Dcal\/$-module
$V\/$ may be extended ``by continuity'' at least to all the
$\hatDcal^{\Ocal}_k\/$ for $k \neq 0\/$.

We shall need the following fact.

\begin{lemma}
  The map $\varphi : \Ocal \rightarrow \Ocal$ given by
  $$
  \varphi
  \left(
  \sum^\infty_{n = 0} f_n z^n
  \right)
  = \sum^\infty_{n = 0} | f_n | z^n
  $$
  is continuous.
\end{lemma}

\Proof
Given $f = \sum^\infty_{n=0} f_n z^n \in \Ocal\/$, where $f_n = |
f_n | e^{i \theta n} \in \C\/$, we let
$$
f^0 ( z ) = \sum^\infty_{n = 0} e^{-i \theta n} z^n, \quad
f^\ast ( z ) = \sum^\infty_{n = 0} | f_n | z^n.
$$
Let $B_R = \{ z \in \C | |z| \leq R \}\/$ denote the disk of
radius $R\/$ and let $C_R\/$ be its boundary.  Note that $f^0 ( z
)\/$ is holomorphic in each $B_{1 -\epsilon}\/$ for $0 < \epsilon
< 1\/$ and that $\max_{B_{1-\epsilon}} | f^0 ( z ) | \leq
\frac1\epsilon\/$.

We need to estimate $| f^\ast ( z ) |\/$ on each disk $B_R\/$.
Take $R_1 > R\/$ and note that for $| w | < R_1\/$ we have:
\begin{equation}
  f^\ast ( w ) = \frac{1}{2 \pi i}
  \int_{C_{R_1}} f ( z ) f^0 \left( \frac{w}{z} \right)
  \frac{dz}{z}.
  \label{eq:wlessthanr}
\end{equation}
{}From \myref{eq:wlessthanr} we see that
$$
\max_{B_R} | f^\ast ( w ) | \leq \max_{B_{R_1}} | f ( z ) |
\cdot \max_{\scriptstyle w \in B_R \atop \scriptstyle | z |
= R_1}
\left| f^0
  \left( \frac{w}{z} \right)
\right| \leq
\frac{1}{1 - R / R_1} \max_{B_{R_1}} | f ( z ) |.\qed
$$

\begin{proposition}
  Let $V$ be a quasifinite $\hatDcal$-module.  Then the action of
  $\hatDcal$ on $V$ naturally extends to the action of
  $\hatDcal^{\Ocal}_k$ on $V$ for any $k \neq 0$.
\end{proposition}

\Proof
Let $V = \bigoplus_p V_p\/$ be the $\Z\/$-gradation of $V\/$,
$\dim V_p < \infty\/$ for all $p\/$.  Consider the space
$$
\Hom ( V, V) = \bigoplus_{p, q} \Hom ( V_p, V_q )
$$
with the topology of direct sum of finite dimensional spaces
$\Hom ( V_p, V_q )\/$.  We can assume that the $V_p\/$ are normed
spaces, and spaces $\Hom ( V_p, V_q )\/$ have induced norms
$\|\/$, $\|_{p, q}\/$.

We will show that map $\hatDcal_k \rightarrow \Hom ( V, V )\/$ for
$k \neq 0\/$ is continuous.  To do this we have to estimate the
norm of the operator $z^k D^n\/$ in the space $\Hom ( V_p, V_{p +
k} )\/$ for fixed $k\/$ and $p\/$ and for arbitrary $n\/$.  We have:
\begin{equation}
  z^k D^n =
  \frac{1}{( 2k )^n}
  \left(
  \ad D^2 - k^2
  \right)^n z^k.
  \label{eq:fixed-k}
\end{equation}
The operator $B = \ad D^2 - k^2 : \Hom ( V_p, V_{p+k} )
\rightarrow \Hom ( V_p, V_{p+k} )\/$ acts between
finite-dimensional normed spaces, hence we obtain from
\myref{eq:fixed-k}:
\begin{equation}
  \left\|
  z^k D^n
  \right\|_{p, p+k} \leq A \cdot \alpha^n,\quad \mbox{where}\quad
  A = \| z^k \|, \quad \alpha = \| B / 2 k \|.
  \label{eq:normed-spaces}
\end{equation}
It follows that
$$
\| z^k f ( D ) \|_{p, p+k} =
\left\|
\sum_{n \geq 0} f_n z^k D^n
\right\|_{p, p+k} \leq
\sum_{n \geq 0} | f_n |\, \| z^k D^n \|_{p, p+k} \leq
A \cdot \sum_{n \geq 0} | f_n | \alpha^n = A \varphi ( f ) ( \alpha ).
$$
Thus, by Lemma 4.3, the map $\hatDcal_k \rightarrow \Hom ( V, V
)\/$ is continuous for $k \neq 0\/$.  Hence this map may be
extended to the completion:  $\hatDcal^\Ocal_k \rightarrow \Hom (
V, V )\/$ (the completion of $\C [ w ]\/$ in topology of
uniform convergence on compact sets is $\Ocal\/$).\qed

\mysubsection
We return now to the $\Z\/$-graded complex Lie algebra $\gfrak^{[m]} :=
\hatgl ( \infty ) [ m ] = \gltilde ( \infty, R_m ) + R_m\/$
introduced in Section 3.3.  Recall that it is a
central extension of the Lie algebra $\gltilde ( \infty, R_m )\/$
over $\C\/$ by the $m + 1\/$-dimensional space $R_m\/$.

An element $\lambda \in ( \gfrak^{[m]}_0 )^\ast\/$ is usually given
by its {\em labels\/}
$$
\lambda^{(j)}_k = \lambda ( t^j E_{kk} ),\quad k \in \Z,\quad j = 0,
\ldots, m,
$$
and {\em central charges\/}
$$
c_j = \lambda ( t^j ), \quad
j = 0, 1, \ldots, m.
$$

Let
\begin{equation}
  h^{(j)}_k = \lambda^{(j)}_k - \lambda^{(j)}_{k+1} + \delta_{k,
  0} c_j, \quad
  k \in \Z,\; j = 0, \ldots, m.
  \label{eq:let}
\end{equation}
As usual, we have the irreducible highest weight
$\gfrak^{[m]}\/$-module $L ( \gfrak^{[m]}, \lambda )\/$
associated to $\lambda\/$.  The proof of the following
proposition is standard:

\begin{proposition}
  The $\gfrak_{[m]}\/$-module $L ( \frak g^{[m]}, \lambda )$ is
  quasifinite if and only if for each $j = 0, 1, \ldots, m$ all
  but finitely many of the $h^{(j)}_k\/$ are zero.
\end{proposition}

\mysubsection
Give $\vec m = ( m_1, \ldots, m_N ) \in \Z^N_+\/$, we let $\frak
g^{[\vec m]} = \hatgl ( \infty ) [ \vec m ] = \bigoplus^N_{i = 1}
\frak g^{[m_i]}\/$.  By Proposition 3.3, we have a surjective
homomorphism of Lie algebras over $\C\/$:
\begin{equation}
  \hat\varphi^{[\vec m]}_{\vec s} =
  \bigoplus^N_{i = 1} \hat\varphi^{[m_i]}_{s_i} :
  \hatDcal \rightarrow \frak g^{[ \vec m ]}.
  \label{eq:lieoverc}
\end{equation}

Choose a quasifinite $\lambda ( i ) \in \Bigl( \frak
g^{[m_i]}_0 \Bigr)^{\ast}\/$ and let $L ( \frak g^{[m_i]},
\lambda ( i ) )\/$ be the corresponding irreducible $\frak
g^{[m_i]}\/$-module.  Then
$$
L \left( \frak g^{[ \vec m ]}, \vec\lambda \right) =
\bigotimes^N_{i = 1} L \left( \frak g^{[m_i]}, \lambda ( i )
\right)
$$
is an irreducible $\frak g^{[\vec m]}\/$-module.  Using the
homomorphism \myref{eq:lieoverc}, we make $L ( \frak g^{[\vec
m]}, \vec\lambda )\/$ a $\hatDcal\/$-module, which we
shall denote by $L^{[\vecm]}_{\vecs} ( \vec\lambda )\/$.

We can prove now the following important Theorem.

\begin{theorem}
  Consider the embedding $\hat\varphi^{[\vec m]}_{\vec s} :
  \hatDcal \rightarrow \hatgl ( \infty ) [ \vec m ]$ (recall
  that $s_i - s_j \not\in \Z$ if $i \neq j$) and let $V$ be a
  quasifinite $\hatgl ( \infty ) [ \vec m ]$-module.  Then any
  $\hatDcal$-submodule of $V$ is a $\hatgl ( \infty ) [ \vec m
  ]$-submodule as well.  In particular, the $\hatDcal$-modules
  $L^{[ \vec m]}_{\vecs} ( \vec\lambda )$ are irreducible.
\end{theorem}

\Proof
Let $U\/$ be a ($\Z\/$-graded) $\hatDcal\/$-submodule of
$V\/$.  $U\/$ is a quasifinite $\hatDcal\/$-module as well,
hence, by Proposition 4.3, it can be extended to
$\hatDcal^\Ocal_k\/$ for any $k \neq 0\/$.  But by Proposition
3.2, the map $\varphi^{[\vec m]}_s : \Dcal^\Ocal_k \rightarrow
\gltilde ( \infty ) [ \vec m ]_k\/$ is surjective for any $k \neq
0\/$.  Thus $U\/$ is invariant with respect to all members of
the principal gradation $\gltilde ( \infty ) [ \vec m ]_k\/$ with
$k \neq 0\/$.  Since $\hatgl ( \infty ) [ \vec m ]\/$ coincides
with its derived algebra, this proves the theorem.\qed

\mysubsection
By Proposition 4.4 and Theorem 4.5, the $\hatDcal\/$-modules
$L^{[ \vec m]}_{\vecs} ( \vec\lambda )\/$  are irreducible
quasifinite highest weight modules.  Using formulas
\myref{eq:taylor1} and \myref{eq:linearmap2}, it is easy to
calculate the generating series $\Delta_{\vecm, \vecs,
\vec\lambda} ( x ) = \sum_{n \geq 0} \Delta_n / n!\/$ of the
highest weight and the central charge $c\/$ of the
$\hatDcal\/$-module $\Lmslambda\/$.
We have
\begin{equation}
  \Delta_{m, s, \lambda} ( x ) =
     -\sum^m_{j = 0} \sum_{i \in \Z}
     \left(
     \lambda^{(j)}_i / j!
     \right)
     x^j e^{(s - i) x} +
     \frac{\displaystyle
       c_0 ( e^{sx} - 1 ) + \sum^m_{j = 1} ( c_j / j! )
          x^j e^{s x}}
       {\displaystyle
          e^x - 1}
   \label{eq:sumsum1}
\end{equation}
\begin{equation}
  c = c_0
  \label{eq:sumsum2}
\end{equation}
and
\begin{equation}
   \Delta_{\vecm, \vecs, \vec\lambda} ( x ) =
   \sum_i \Delta_{m_i, s_i, \lambda (i)} ( x ), c =
   \sum_i c_0 ( i ).
  \label{eq:sumsum3}
\end{equation}
Introduce the polynomials (see \myref{eq:let}):
\begin{equation}
  g_k ( x ) =
  \sum^m_{j = 0} h^{(j)}_k x^j / j!\quad ( k \in \Z ).
  \label{eq:polynomials}
\end{equation}
Then \myref{eq:sumsum1} can be rewritten as follows:
\begin{equation}
  \Delta_{m, s, \lambda} ( x ) =
  \frac{\displaystyle\sum_{k \in \Z} e^{( s - k ) x} g_k (
  x ) - c_0}{e^x - 1}.
  \label{eq:follows}
\end{equation}
Using these formulas, it is not difficult to see that any
irreducible quasifinite highest weight module $L ( \hatDcal,
\lambda )\/$ can be thus obtained in an essentially unique way.
More precisely, we have the following proposition.

\begin{theorem}
  Let $L = L ( \hatDcal, \lambda )$ be an irreducible quasifinite
  highest weight module
  with central charge $c$ and $\Delta_{\lambda} ( x ) =
  \phi ( x ) / ( e^x - 1 )$, where $\phi ( x )$ is a
  quasipolynomial such that $\phi ( 0 ) = 0$ (see Theorem 4.2).
  We write $\phi ( x ) + c = \sum_{s \in \C} p_s ( x ) e^{s
  x}$, where $p_s ( x )$ are polynomials.  We decompose the
  set $\{ s \in \C| p_s ( x ) \neq 0 \}$ in a disjoint union of
  congruence classes ${\rm mod}\, \Z$.  Let $S = \{ s, s+k_1,$
  \newline
  $s + k_2,
  \ldots \}$ be such a congruence class, let $m = \max_{s \in
  S} \deg p_s$ and let $h^{(j)}_{k_r} = \left( \frac{d}{dx}
  \right)^j p_{s + k_r} ( 0 )$.  We associate to $S$ the
  $\gltilde ( \infty ) [ m ]$-module $L^{[m]} ( \lambda_S )$
  with the central charges
  \begin{equation}
    c_j = \sum_{k_r} h^{(j)}_{k_r},
    \label{eq:charges}
  \end{equation}
and labels
\begin{equation}
  \lambda^{(j)}_i = \sum_{k_r \geq i} \tilde{h}^{(j)}_{k_r},
  \label{eq:labels}
\end{equation}
where $\tilde{h}^{(j)}_k = h^{(j)}_k - c_j \delta_{k, 0}\/$.
Then the $\hatDcal\/$-module $L\/$ is isomorphic to the tensor
product of all the modules $L^{[m]}_s ( \lambda_S )\/$.
\end{theorem}

\Proof
The tensor product $L'\/$ of all the modules $L^{[m]}_s (
\lambda_S )\/$ is irreducible due to Theorem 4.5.  It remains to
show that $L'\/$ has the same highest weight as $L\/$ does.  This
is done by exploiting the observation (used already before) that
$- \Delta ( x )\/$ is the value of the highest weight of $L'\/$ on
$e^{x D}\/$, and using the formulas (\ref{eq:sumsum2}--5).\qed

\Remark
Changing the representative $s\/$ in $S\/$ amounts to the shift
$\nu^j\/$ of $gl ( \infty ) [ m ]\/$.  Up to these shifts the
above construction of $L\/$ via the embedding $\hatDcal
\rightarrow gl ( \infty ) [ \vecm ]\/$ is unique.

\mysection{Unitary quasifinite highest weight modules over
$\hatDcal\/$.}

\mysubsection
It is easy to see that any anti-involution $\omega\/$ of the
associative algebra $\Dcal^a\/$, such that $\omega ( \Dcal^a_j )
= \Dcal^a_{-j}\/$ and $\omega ( D ) = D\/$, by a rescaling of $z
\/$ can be brought to the following form:
\begin{equation}
  \omega ( z^k f ( D ) ) =
  \bar{f} ( D ) z^{-k} = z^{-k} \bar{f} ( D - k ),
  \label{eq:rescale}
\end{equation}
where for $f ( D ) = \sum_i f_i D^i\/$ we let $\bar{f} ( D ) =
\sum_i \bar{f}_i D^i\/$.  The involution $\omega\/$ given by
\myref{eq:rescale} extends to the whole algebra
$\Dcal^{a\Ocal}\/$.

Note that
\begin{equation}
  \Psi ( \omega ( A ), \omega ( B ) ) =
  \Psi ( B, A), \qquad
  A, B \in \Dcal^\Ocal.
  \label{eq:something}
\end{equation}
Hence the anti-involution $\omega\/$ of the Lie algebras
$\Dcal\/$ and $\Dcal^\Ocal\/$ lifts to an anti-involution of
their central extensions $\hatDcal\/$ and $\hatDcal^a\/$, such
that $\omega ( C ) = C\/$, which we again denote by $\omega\/$.

\Remark (a) The Virasoro subalgebra $\Vir ( \beta )\/$ (defined
by \myref{eq:virasoro1}) is $\omega\/$-stable if an only if $\beta
= \frac12\/$.

(b) Under the homomorphism $\varphi_s = \varphi_{s, 0} : \Dcal^{a
\Ocal} \rightarrow \Mtilde ( \infty, \C )\/$ we have
$$
\left(
  \varphi_s ( z^k f ( D ) )
\right)^\ast = \varphi_{\bar s}
\left(
  \omega ( z^i f ( D ) )
\right).
$$
Here $A^\ast\/$ stands for the complex conjugate transpose
of the matrix $A \in \Mtilde ( \infty, \C )\/$.

(c) (see e.g.\ \cite{K}) For the involution $\omega\/$ of $\hatgl
( \infty, \C )\/$ defined by $\omega ( A ) = {}^t\! \bar{A}\/$,
$\omega ( 1 ) = 1\/$, a highest weight $\hatgl ( \infty, \C
)\/$-module with highest weight $\lambda\/$ and central change
$c\/$ is unitary if and only if the numbers $h_i^{( 0 )}\/$ (see
\myref{eq:let}) are non-negative integers and $c = \sum_i h_i^{(
0 )}\/$.

\mysubsection
In this section we shall classify all unitary (irreducible)
quasifinite highest weight modules over the Lie algebra
$\hatDcal\/$ with respect to the anti-involution $\omega\/$.

\begin{lemma}
  Lev $V$ be a unitary quasifinite highest weight
  module over $\hatDcal$ and let $b ( w )$ be its
  characteristic polynomial.  Then $b ( w )$ has
  only simple real roots.
\end{lemma}

\Proof
Let $v_\lambda\/$ be a highest weight vector of $V\/$ and let $\Delta_j
= \lambda ( D^j ) \in \R\/$ be the labels of $\lambda\/$.
Consider the element $S = -\frac12 ( D^2 - \Delta_2 - 1 ) \in
\hatDcal\/$.  It is easy to check that for any $j \in \Z_+\/$ we
have:
\begin{equation}
  S^j ( z^{-1} v_\lambda ) = (z^{-1} D^j ) v_\lambda.
  \label{eq:jinzplus}
\end{equation}
By definition of the characteristic polynomial we have:
\begin{equation}
  \Bigl( z^{-1} b ( D ) \Bigr)
  v_\lambda = 0,
  \label{eq:firstchar1}
\end{equation}
\begin{equation}
  \left\{
  ( z^{-1} D^j ) v_\lambda  | 0 \leq j < n
  \right\}\quad
  \mbox{is a basis of}\quad
  V_{-1},
\label{eq:firstchar2}
\end{equation}
where $n = \deg b ( w )\/$.  It follows from \myref{eq:jinzplus}
and \myref{eq:firstchar1} that
\begin{equation}
  b ( S )
  ( z^{-1} v_\lambda ) = 0,
  \label{eq:firstchar3}
\end{equation}
and it follows from \myref{eq:firstchar2} that
\begin{equation}
  \left\{
  S^j ( z^{-1} v_\lambda ) |
  0 \leq j < n
  \right\} \quad
  \mbox{is a basis of}\quad
  V_{-1}.
  \label{eq:firstchar4}
\end{equation}
We conclude from \myref{eq:firstchar3} and \myref{eq:firstchar4}
that $b ( w )\/$ is the characteristic polynomial of the operator
$S\/$ on $V_{-1}\/$. Operator $S$ is selfadjoint, hence roots of
$b ( w )\/$ are real.

Let $\mu\/$ be a root of $b ( w )\/$ of multiplicity $m\/$, so
that $b ( w ) = c ( w ) ( w - \mu )^m\/$, $c(w) \in \C [ w ]\/$.
Then
$$
v := ( S - \mu )^{m - 1} c ( S ) ( z^{-1} v )
$$
is a non-zero vector in $V_{-1}\/$, but
$$
h ( v, v) = h ( c ( S ) ( z^{-1} v ), ( S - \mu )^{2 m - 2} c ( S
) ( z^{-1} v ) ) = 0 \quad \mbox{if}\quad m \geq 2,
$$
by \myref{eq:firstchar3}.  Hence the unitarity forces $m =
1\/$.\qed

\begin{theorem}
  (a) A quasifinite irreducible highest weight module
  $L ( \hatDcal, \lambda )$ is unitary if and only if
  \begin{equation}
    \Delta_\lambda ( x ) = \sum_i n_i
    \frac{e^{s_i x} - 1}{e^x - 1}
    \label{eq:unitary1}
  \end{equation}
for some positive integers $n_i\/$ and real numbers $s_i\/$, such
that
\begin{equation}
  c = \sum_i n_i.
  \label{eq:posintegers}
\end{equation}

(b) Any unitary quasifinite $\hatDcal\/$-module is obtained by
taking tensor product of $N\/$ unitary irreducible quasifinite
highest weight modules over $\hatgl ( \infty, \C )\/$ and
restricting to $\hatDcal\/$ via an embedding
$\hat\varphi^{[0]}_{\vecs}\/$, where $\vecs = (s_1, \ldots, s_N
)\/$ is a real vector and $s_i - s_j \not\in \Z\/$ if $i \neq
j\/$.
\end{theorem}

\Proof
By Proposition 4.6, being a quasifinite irreducible highest
weight $\hatDcal\/$-module, $V\/$ is isomorphic to one of the
modules $\Lmslambda\/$.  It follows from Lemma 5.2 and Corollary
4.2 that $\vecm = 0\/$.  Now the claim (b) follows from Remarks
5.1(b) and (c).  The claim (a) follows from (b) and
(\ref{eq:sumsum1} and 2).\qed

\begin{corollary}
  Suppose that only finitely many labels $\Delta_n$ of $\lambda$
  are non-zero.  Then the $\hatDcal$-module $L ( \lambda )$ is
  unitary if and only if
  \begin{equation}
    c = \Delta_0 \in \Z_+\quad \mbox{and}\quad
    \Delta_j = 0\; \mbox{for}\; j > 0.
    \label{eq:unitary}
  \end{equation}
\end{corollary}

\Proof
By the hypothesis, $\Delta_\lambda ( x )\/$ is a polynomial of
degree $N\/$, where $N = \max_L \{ n | \Delta_n \neq 0 \}\/$.  By
Corollary 4.2, it follows that the characteristic
polynomial of $L ( \lambda )\/$ is $w^{N+1}\/$.  Hence, by Lemma
5.2, $N = 0\/$, i.e., $\Delta_j = 0\/$ for $j > 0\/$.  The rest
of \myref{eq:unitary} follows from Theorem 5.2\qed

\Remark
The $\hatDcal\/$-modules of Corollary 5.2 are obtained by taking
the embedding $\hat\varphi^{[0]}_0 : \hatDcal \rightarrow \hatgl
( \infty, \C )\/$ and composing it with the irreducible highest
weight $\hatgl ( \infty, \C )\/$-module with a non-negative
integral central charge and zero labels.

\Example
Consider the following parabolic subalgebra of the Lie algebra
$\gfrak = \hatgl ( \infty , \C )\/$:
$$
\pfrak =
\left\{
 ( a_{ij} )_{i, j \in \C} + \C C |
 a_{ij} = 0 \; \mbox{if}\; i > 0 \geq j
\right\}
$$
and let $F_0 = E_{10}\/$.  Given $c \in \C\/$, denote by $M_c\/$
the generalized Verma module $M ( \gfrak, \pfrak, \lambda_0 )\/$
where $\lambda_0\/$ is the highest weight such that $\lambda_0 (
E_{jj} ) = 0\/$ for all $j\/$ and $\lambda_0 ( C ) = c\/$.  Then
we have
$$
\begin{array}{rcl}
  L ( \gfrak, \lambda_0 ) & = &
    M_c\; \mbox{if}\; c \not\in \Z_+,\\
  L ( \gfrak, \lambda_0 ) & = &
    M_c / \Ucal ( \frak g ) ( F^{c+1}_0 v_{\lambda_0} )\; \mbox{if}\; c
    \in \Z_+.
\end{array}
$$
Consider the homomorphism $\hat\varphi_0 : \hatDcal \rightarrow
\gfrak\/$ given by:
$$
\hat\varphi_0 ( z^k f ( D ) ) =
\sum_{j \in \Z} f ( -j ) E_{j-k, j},\; \hat\varphi_0 ( C ) = 1.
$$
When restricted to $\hatDcal\/$, the module $L ( \frak g,
\lambda_0 )\/$ remains an irreducible quasifinite highest weight
module with zero labels and central charge $c\/$.  The singular
vector $F^{c+1}_0 v_{\lambda_0}\/$ of the $\frak g\/$-module
$M_c\/$ remains singular for the $\hatDcal\/$-module $M_c\/$.  It
is a multiple of the following vector:
$$
\left(
z^{-1} \prod^c_{s=1} ( D^2 - s^2 )
\right)^{c+1} v_{\lambda_0}.
$$

\mysection{Quasifinite highest weight modules over quantum
pseudo-differential operators.}

\mysubsection
The $q\/$-analogue of the algebra $\Dcal^a\/$ is the algebra
$\Dcalaq\/$ of all regular difference operators on $\C^\times\/$
(see Section 1.4).  However, a more important algebra is the
algebra of quantum pseudo-differential operators $\Scal^a_q\/$ (which
contains $\Dcalaq\/$ as a subalgebra).  This associative algebra
is obtained by the construction explained in Section 1.1 by
taking the algebra $A = \C [ w, w^{-1} ]\/$ and its automorphism
$\sigma\/$ defined by $\sigma ( w ) = q w\/$, where $q \in
\C^\times\/$:
$$
\Scal^a_q = A_\sigma [ z, z^{-1} ].
$$
Explicitly, let $T_q\/$ denote the following operator on $\C [ z,
z^{-1} ]\/$, where $q \in \C^\times\/$:
$$
T_q f ( z ) = f ( q z ).
$$
Then $\Scal^a_q\/$ is the associative algebra of all operators on $\C
[ z, z^{-1} ]\/$ of the form
$$
E = \sum_{k \in \Z} e_k ( z ) T^k_q,\quad
\mbox{where}\quad
e_k ( z ) \in \C [ z, z^{-1} ]\; \mbox{and sum is finite}.
$$
As before, we write such an operator as a linear combination of
operators of the form $z^k f ( T_q )\/$, where $f\/$ is a Laurent
polynomial in $T_q\/$.  Then the product is given by
\begin{equation}
  \left(
  z^m f ( T_q )
  \right) ( z^k g ( T_q ) ) =
  z^{m + k} f ( q^k T_q ) g ( T_q ).
  \label{eq:product}
\end{equation}

Let $S_q\/$ denote the Lie algebra obtained from $\Scal^a_q\/$ by
taking the usual bracket.  Let $\Scal'_q = [ \Scal_q, \Scal_q
]\/$.  We have:
$$
\Scal_q = \Scal'_q \oplus \C T^0_q\quad \mbox{(direct sum of ideals)}.
$$
Thus, representation theory of $\Scal_q\/$ reduces to that of
$\Scal'_q\/$.

Taking the trace form $\tr_0 ( \sum_j c_j w^j ) = c_0\/$, we
obtain by the general construction of Section 1.3 the following
2-cocycle on $\Scal_q\/$:
\begin{equation}
  \Psi ( z^m f ( T_q ),
  z^k g ( T_q ) ) =
  m \delta_{m, -k} \tr_0 f ( q^{-m} w ) g ( w ).
  \label{eq:genconstr}
\end{equation}
The associated central extension of $\Scal'_q\/$ is denoted by
$\hat\Scal_q = \Scal'_q + \C C\/$.  As we have mentioned in Remark
1.6(b), this is a well-known Lie algebra studied by many authors.

We will show that the representation theory of the Lie algebra
$\hat\Scal_q\/$ with $|q| \neq 1$ is quite similar to that of
$\hatDcal\/$.  Details of most of the proofs will be omitted,
being similar as well.

\mysubsection
Let $\Ocal\/$ denote (in this section) the algebra of all
holomorphic functions in $\C^\times = \C \backslash \{ 0 \}\/$.
We define a completion $\Scal^{a\Ocal}_q\/$ of the algebra
$\Scal^a_q\/$ by considering operators of the form $z^k f ( T_q
)\/$, where $f \in \Ocal\/$.  We extend the product
\myref{eq:product} to $\Scal^{a\Ocal}_q\/$ and denote by
$\Scal^\Ocal_q\/$ the corresponding Lie algebra.  The cocycle
\myref{eq:genconstr} extends to $\Scal^\Ocal_q\/$ and we let
$\hat \Scal^\Ocal_q = \Scal^{\Ocal'}_q + {\cal C} C\/$ be the
corresponding central extension.

Consider the associative algebra $R_m = \C [ t ] / ( t^{m+1} )\/$
and let $s \in \C\/$.  Then we have the following embedding
$\varphi_{s, t} : \Scal^a_q \rightarrow \Mtilde ( \infty, R_m
)\/$ of $\Z\/$-graded associative algebras over $\C\/$ (cf.\
\cite{GL}):
\begin{equation}
  \varphi^{[m]}_s ( z^k f ( T_q ) ) =
  \sum_{j \in \Z} f ( sq^{-j + t} ) E_{j - k, j}.
  \label{eq:associative}
\end{equation}
It extends to a homomorphism $\varphi^{[m]}_s : \Scal^{a\Ocal}_q
\rightarrow \Mtilde ( \infty, R )\/$.

\begin{lemma}
  The homomorphisms $\varphi^{[m]}_s$ are surjective provided
  that $| q | \neq 1$.
\end{lemma}

Let $\varphi_s = \varphi^{[0]}_s : \Scal^{a\Ocal}_q \rightarrow
\Mtilde ( \infty, \C )\/$.  We have (cf.\ Section 3.4):
\begin{equation}
  \varphi_{qs} ( z^k f ( T_q ) ) =
  \nu \varphi_s ( z^k f ( T_q ) ).
  \label{eq:have}
\end{equation}
We define a {\em quantum monodromic loop\/} to be a holomorphic
map $f : \C^\times \rightarrow \Mtilde ( \infty, \C )\/$ such
that $f ( qw ) = \nu f ( w )\/$.  Denote by ${\cal L}_{q, \nu}
\Mtilde ( \infty )\/$ the associative algebra of all monodromic
loops.  Then we have an isomorphism
\begin{equation}
  \varphi : \Scal^{a\Ocal}_q \simrightarrow {\cal L}_{q, \nu}
  \Mtilde ( \infty )
  \label{eq:isomorphism}
\end{equation}
defined by the same formula as \myref{eq:linearmap}.  Note that
the quantum monodromic loops are sections of the vector bundle on
the torus $\C^\times / \{ q^n | n \in \Z \}\/$ (with modular
parameter $( \log q ) / 2 \pi i\/$) with fiber $\Mtilde ( \infty,
\C )\/$ and transition function $\nu\/$ in a small neighbourhood
of the circle $| w | = | q |\/$.

Denote by ${\cal L}_{q, \nu} \gltilde ( \infty )\/$ the Lie
algebra obtained from ${\cal L}_{q, \nu} \Mtilde ( \infty )\/$ by
taking the usual bracket.  Considering the Laurent expansion at
0:
$$
C ( f ( w ) , g ( w ) ) =
\sum_{n \in \Z} C_n ( f, g ) w^n,
$$
we obtain $\C\/$-valued 2-cocycles $C_n\/$ on this Lie algebra.
Denote by ${\cal L}_{q,\nu} \gltilde ( \infty )^\wedge\/$ the
corresponding to $C_0\/$ central extension.  Then the isomorphism
$\varphi : S^\Ocal_q \simrightarrow {\cal L}_{q, \nu} \gltilde (
\infty )\/$ lifts to the isomorphism $\hat\varphi : \hat
\Scal^\Ocal_q \simrightarrow {\cal L}_{q, \nu} \gltilde ( \infty
)^\wedge\/$.

\mysubsection
Let $\frak p = \bigoplus_{j \in \Z} \frak p_j\/$ be a parabolic
subalgebra of the Lie algebra $\hat \Scal_q\/$ (i.e., $\frak p_j
= ( \hat\Scal_q )_j\/$ for $j \geq 0\/$ and $( \hat\Scal_q)_j
\neq 0\/$ for some $j < 0\/$).  Then for each positive integer
$k\/$ we have $\frak p_{-k} = z^{-k} I_{-k}\/$, where $I_{-k}\/$
is a non-zero ideal of $\C [ w, w^{-1} ]\/$.  Let $b_k ( w )\/$
be the monic polynomial with $b_k ( 0 ) \neq 0\/$ which is a
generator of the ideal $I_{-k}\/$.  The polynomials $b_k\/$, $k =
1, 2, \ldots \/$ are called the characteristic polynomials of
$\frak p\/$.

Given a monic polynomial $b ( w )\/$ with $b ( 0 ) \neq 0\/$, we
let
$$
b^{\rm min}_{k} ( w ) = b ( w ) b ( q^{-1} w ) \cdots
b ( q^{-k+1} w ), \quad
b^{\rm max}_k = {\rm lcm}\, \{ b ( w ), b ( q^{-1} w ), \ldots,
b ( q^{-k+1} w ) \}.
$$
There exist unique parabolic subalgebras of $\hat\Scal_q\/$, which
we denote by $\frak p_{\rm min} ( b )\/$ and $\frak p_{\rm max} (
b )\/$, for which the characteristic polynomials are $\{ b^{\rm
min}_k ( w ) \}\/$ and $\{ b^{\rm max}_k ( w )\}\/$ respectively.
We have
$$
\dim ( \hat S_q )_{-k} / \frak p_{\rm min} ( b )_{-k} =
k \deg b ( w ),
$$
and an analogue of Proposition 2.4 holds verbatim.

Also, we have for a parabolic subalgebra $\frak p\/$ with the
first characteristic polynomial $b ( w )\/$:
$$
  [ \frak p, \frak p ] =
  \left(
  \bigoplus_{k \neq 0} \frak p_k
  \right) \bigoplus ( \hat S_q )^b_0,
$$
where
$$
  ( \hat\Scal_q )^b_0 =
  \left\{
    b ( T_q ) g ( T_q ) - b ( q T_q ) g ( q T_q ) +
    ( \tr_0 b ( w ) g ( w ) ) C | g ( w ) \in \C [ w, w^{-1} ]
  \right\}.
$$

\mysubsection
All the results of Section 4.1 hold for the Lie algebra
$\hat{S}_q\/$ verbatim.  However, the generating series
$\Delta_\lambda ( x )\/$ is defined differently.

We shall characterize $\lambda \in ( \hat\Scal_q )^\ast_0\/$ by
labels $\Delta_n = \lambda ( T^n_q )\/$ ($n \neq 0\/$) and
central charge $c = \lambda ( C )\/$.  Introduce the generating
series
$$
\Delta_\lambda ( x ) = \sum_{\scriptstyle n \in \Z \atop
\scriptstyle n \neq 0} \Delta_n x^{-n}.
$$

\begin{theorem}
  (a) An irreducible highest weight module $L ( \hat S_q, \lambda
  )$ is quasifinite if and only if one of the following
  equivalent conditions holds:
  {\renewcommand{\labelenumi}{(\roman{enumi})}
  \begin{enumerate}
  \item there exists a non-zero polynomial $b ( x )$ such that
    \begin{equation}
      b ( x ) ( \Delta_\lambda ( x ) - \Delta_\lambda ( q^{-1} x
      ) + c ) = 0
      \label{eq:nonzero}
    \end{equation}
  \item there exists a quasipolynomial $P ( x )$ such that
    $$
    ( 1 - q^n ) \Delta_n = P ( n )\; \mbox{for}\; n \neq 0\;
    \mbox{and}\; c = P ( 0 ).
    $$
  \end{enumerate}}

  (b) The monic polynomial of minimal degree satisfying
  \myref{eq:nonzero} is the characteristic polynomial of a
  quasifinite module $L ( \hat S_q, \lambda )$.
\end{theorem}

\Proof
According to Section 6.3, $L ( \hat\Scal_q, \lambda )\/$ is
quasifinite if and only if there exists a non-zero polynomial $b
( w )\/$ such that
$$
\lambda ( g ( T_q ) b ( T_q ) - g ( q T_q )
b ( q T_q ) ) +
c\, \tr_0 ( g ( w ) b ( w ) ) = 0
$$
for each $g ( w ) \in \C [ w, w^{-1} ]\/$.  Taking $g ( w ) =
w^n\/$, and letting $b ( w ) = \sum_j f_j w^j\/$, this can be
rewritten as follows:
$$
\sum_j f_j \Delta_{n+j} ( 1 - q^{n+j} ) +
f_{-n} c = 0\; \mbox{for all}\; n \in \Z.
$$
Multiplying both sides of this equality by $x^{-n}\/$ and summing
over $n \in \Z\/$, we obtain \myref{eq:nonzero}.

The equivalence of (i) and (ii), as well as (b) are clear.\qed

\mysubsection
Choose a branch of $\log q\/$. Let $\tau = (\log q)/(2\pi i)\/ $.
Then any $s \in \C\/$ is uniquely
written as $s = q^a\/$, $a \in \C/\tau^{-1}\Z \/$.  The homomorphism
$\varphi^{[m]}_s : \Scal'_q \rightarrow \gltilde ( \infty ) [ m ]\/$
defined by \myref{eq:associative} lifts to a homomorphism $\hat
\Scal_q \rightarrow \hatgl ( \infty ) [ m ]\/$ of central extensions,
denoted by $\hat\varphi^{[m]}_a\/$, by
$$
\hat\varphi^{[m]}_a \left|_{(\hat \Scal_q )_j} = \varphi^{[m]}_a
\right|_{(\Scal_q)_j}\quad \mbox{if}\quad j \neq 0,
$$
$$
\hat\varphi^{[m]}_a ( T^n_q ) =
\sum_{r \in \Z} q^{(a-r)} E_{rr} +
\frac{q^{an}}{1 - q^n} \sum^m_{j = 0}
( n \log q )^j t^j / j!\quad ( n \neq 0),
$$
$$
\hat\varphi^{[m]}_a ( C ) = 1 \in R_m.
$$
We have results similar to Theorems 4.5, 4.6, and 5.2:

\begin{theorem}
  Assume that $|q| \neq 1$.  Consider the embedding
  $\hat\varphi^{[ \vec m]}_{\vec a} : \hat \Scal_q \rightarrow \hatgl
  ( \infty ) [ \vec m ]$, where $a_i - a_j \not\in \Z + \tau^{-1} \Z $
  if $i \neq
  j$.  Denote the quasifinite $\hatgl ( \infty ) [ \vec m
  ]$-module $L^{[\vec m]} ( \vec\lambda )$, viewed as a $\hat
  \Scal_q$-module via this embedding, by $L^{[\vec m]}_{\vec a} (
  \vec\lambda )$.

  (a) If $V$ is a quasifinite $\hatgl ( \infty ) [ \vec m
  ]$-module, then any submodule of the module $V$, viewed as a
  $\hat \Scal_q$-module via the embedding $\hat\varphi^{[ \vec m
  ]}_{\vec a}$, is a $\hatgl ( \infty ) [ \vec m ]$ submodule as
  well.  In particular, the $\hat \Scal_q$-modules $L^{[ \vec
  m]}_{\vec a} ( \vec\lambda )$ are irreducible.

  (b) Any irreducible quasifinite highest weight module over
  $\hat \Scal_q$ is isomorphic to one of the modules $L^{[\vec
  m]}_{\vec a} ( \vec\lambda )$.

  (c) Let $q\in \R$ , and let $\omega$
  be the anti-involution of $\hat \Scal_q$ defined
  by
  $$
  \omega ( z^k f ( T_q ) ) = z^{-k} \bar f ( q^{-k} T_q ).
  $$
  Then a quasifinite highest weight module over $\hat \Scal_q$ is
  unitary with respect to $\omega$ if and only if
  $$
  \Delta_n = \sum_j
  \frac{n_j q^{a_j n}}{1 - q^n}\quad
  \mbox{(finite sum) for all}\; n \neq 0,
  $$
  where the $n_j$ are positive integers and the $a_j$ are real
  numbers.  (In particular, unitarity implies that $c = \sum_j
  n_j \in \Z_+$.)  Any unitary quasifinite $\hat \Scal_q$-module is
  obtained by taking tensor product of $N$ unitary irreducible
  quasifinite highest weight modules over $\hatgl ( \infty, \C )$
  and restricting to $\hat \Scal_q$ via an embedding $\hat\varphi^{[
  0 ]}_{\vec a}$, $\vec a \in \Bbb R^N$.
\end{theorem}

\Remarks
(a) The labels and the central charge of the $\hat
\Scal_q\/$-module $L^{[m]}_a ( \lambda )\/$ are given by the
following formulas (see \myref{eq:polynomials}):
$$
  \Delta_n = \sum_{k \in \Z} q^{(a - k)n} g_k ( n \log q ) / ( 1
  - q^n ), \quad n \neq 0;\; c = c_0.
$$

(b) A vertex operator construction of unitary quasifinite highest
weight modules over $\hat\Scal_q\/$ with $c = 1\/$ is given in
\cite{GL}.

\mysection{Matrix case}

Let us consider the Lie algebras $M_n \Dcal = \Mat_n ( \Dcal )\/$
and $M_n \Scal_q\/$.  They are twisted Laurent polynomial
algebras with $A = \Mat_n [ w ]\/$, $\sigma ( w ) = w + 1\/$ and
$A = \Mat_n [ w, w^{-1} ]\/$, $\sigma ( w ) = q w\/$
respectively.  The canonical central extension $M_n \hat\Scal_q\/$
is defined via \myref{eq:cocycle} with respect to the trace
functional $\tr_0 : \Mat_n [ w, w^{-1} ] \rightarrow \C \/$
defined by
$$
\tr_0
\left(
\sum_i m_i w^i
\right) = \tr\, m_0 .
$$
Restriction of $\tr_0\/$ to $\Mat_n [ w ]\/$ gives rise to the
canonical central extension $M_n \Dcal^\wedge\/$.

The isomorphism $\C^n [ z, z^{-1} ] \simrightarrow \C [ z, z^{-1}
]\/$ defined by
$$
e_i z^j \rightarrow z^{j n + i}
$$
defines the isomorphism $\Mat_n ( \Mtilde ( \infty ) )
\simrightarrow \Mtilde ( \infty )\/$.  Combined with isomorphisms
$\hatDcal^{\Ocal} \simrightarrow \Lcal_\nu \gltilde ( \infty
)^\wedge\/$  and $\Scal^{\Ocal\wedge}_q \simrightarrow \Lcal_{q,
\nu} \gltilde ( \infty )^\wedge\/$ it gives Lie algebra
isomorphisms:
$$
\begin{array}{rcccl}
M_n \Dcal^{\Ocal\wedge} & \simrightarrow &
  \Lcal_\nu \Mat_n \gltilde
  ( \infty )^\wedge & \simrightarrow & \Lcal_{\nu^n}
  \gltilde ( \infty )^\wedge\\
M_n \Scal^\wedge_q & \simrightarrow & \Lcal_{q, \nu} \Mat_n
  \gltilde ( \infty )^\wedge & \simrightarrow & \Lcal_{q, \nu^n}
  \gltilde ( \infty )^\wedge.
\end{array}
$$

The representation theory of the Lie algebras $M_n
\Dcal^\wedge\/$ and $M_n \Scal^\wedge_q\/$ is similar to that of
$\hatDcal\/$ and $\hat\Scal_q\/$.  All irreducible quasifinite
highest weight modules over $M_n \Dcal^\wedge\/$ and $M_n
\Scal^\wedge_q\/$ are constructed by embedding in $\hatgl (
\infty ) [ \vec m ]\/$ and restricting an irreducible quasifinite
highest weight module over the latter.

An anti-involution of an algebra $B\/$ combined with the matrix
transposition defines an anti-involution of $\Mat_n ( B )\/$.
All quasifinite unitary highest weight modules over $M_n
\Dcal^\wedge\/$ and $M_n \Scal^\wedge_q\/$ are modules over $gl (
\infty ) [ \vec 0 ]\/$.  In particular, unitary modules over $M_n
\Dcal^\wedge\/$ and $M_n \Scal^\wedge_q\/$ have positive integer
central charge.

\bibliographystyle{alpha}

\end{document}